\documentclass[journal]{IEEEtran}
\usepackage{cite}

\ifCLASSINFOpdf
   \usepackage[pdftex]{graphicx}
  \graphicspath{{Figures/}}
\else
   \usepackage[dvips]{graphicx}
\fi
\usepackage{amsmath}
\usepackage{amsfonts}

\usepackage{algpseudocode}
\usepackage{algorithm}

\algrenewcommand\algorithmicrequire{\textbf{Input:}}
\algrenewcommand\algorithmicensure{\textbf{Output:}}

\algnewcommand{\Inputs}[1]{%
  \State \textbf{Inputs:}
  \Statex \hspace*{\algorithmicindent}\parbox[t]{.8\linewidth}{\raggedright #1}
}
\algnewcommand{\Init}[1]{%
  \State \textbf{Initialization:}
  \Statex \hspace*{\algorithmicindent}\parbox[t]{.8\linewidth}{\raggedright #1}
}
\algnewcommand{\Outputs}[1]{%
  \State \textbf{Output:}
  \Statex \hspace*{\algorithmicindent}\parbox[t]{.8\linewidth}{\raggedright #1}
}

\usepackage{array}

\ifCLASSOPTIONcompsoc
 \usepackage[caption=false,font=normalsize,labelfont=sf,textfont=sf]{subfig}
\else
 \usepackage[caption=false,font=footnotesize]{subfig}
\fi
\usepackage{url}
\usepackage[backref]{hyperref} 
\makeatletter
\def\UrlAlphabet{%
      \do\a\do\b\do\c\do\d\do\e\do\f\do\g\do\h\do\i\do\j%
      \do\k\do\l\do\m\do\n\do\o\do\p\do\q\do\r\do\s\do\t%
      \do\u\do\v\do\w\do\x\do\y\do\z\do\A\do\B\do\C\do\D%
      \do\E\do\F\do\G\do\H\do\I\do\J\do\K\do\L\do\M\do\N%
      \do\O\do\P\do\Q\do\R\do\S\do\T\do\U\do\V\do\W\do\X%
      \do\Y\do\Z}
\def\UrlDigits{\do\1\do\2\do\3\do\4\do\5\do\6\do\7\do\8\do\9\do\0}
\g@addto@macro{\UrlBreaks}{\UrlOrds}
\g@addto@macro{\UrlBreaks}{\UrlAlphabet}
\g@addto@macro{\UrlBreaks}{\UrlDigits}
\makeatother

\hypersetup{
hidelinks,
colorlinks=true,
linkcolor=black,
citecolor=black
}

\usepackage{icomma}
\usepackage{comment}

\DeclareMathOperator{\diag}{diag}

\def\mbf#1{\mathbf{#1}}
\def\mbb#1{\mathbb{#1}}

\def\mcal#1{\mathcal{#1}}
\def\bs#1{\boldsymbol{#1}}
\def\myx{\mbf{x}}

\def\myz{\mbf{z}}
\def\mys{\mbf{s}}
\def\mytheta{\bs{\theta}}
\def\myphi{\bs{\phi}}
\def\myvarphi{\bs{\varphi}}

\newcommand{\appropto}{\mathrel{\vcenter{
  \offinterlineskip\halign{\hfil$##$\cr
    \propto\cr\noalign{\kern2pt}\sim\cr\noalign{\kern-2pt}}}}}

\usepackage{subfig}

\usepackage{booktabs}

\usepackage{inputenc}
\usepackage[table]{xcolor}
\usepackage{bbding}

\newcommand{\etal}{\textit{et al.}}

\begin{document}

%
\title{Unsupervised Speech Enhancement using \\  Dynamical Variational Autoencoders}
%
%
%

\author{Xiaoyu~Bie,$^1$ Simon Leglaive,$^2$~\IEEEmembership{Member,~IEEE}, Xavier Alameda-Pineda,$^1$~\IEEEmembership{Senior Member,~IEEE}, Laurent Girin$^3$
\thanks{$^1$ Inria Grenoble Rh\^one-Alpes, Univ. Grenoble-Alpes, France}
\thanks{$^2$ CentraleSup\'elec, IETR (UMR CNRS 6164), France}
\thanks{$^3$ Univ. Grenoble Alpes, Grenoble-INP, CNRS, GIPSA-lab, France}
\thanks{This research was supported by ANR-3IA MIAI (ANR-19-P3IA-0003), ANR-JCJC ML3RI (ANR-19-CE33-0008-01), H2020 SPRING (funded by EC under GA \#871245). This work was performed using HPC resources from the “Mésocentre” computing center of CentraleSupélec and École Normale Supérieure Paris-Saclay supported by CNRS and Région Île-de-France.}
}

%
%

\markboth{Submitted to IEEE TASLP,~Vol.~XX, No.~YY, 2022}%
{Shell \MakeLowercase{\textit{et al.}}: Bare Demo of IEEEtran.cls for IEEE Journals}
%



\maketitle





\begin{abstract}
Dynamical variational autoencoders (DVAEs) are a class of deep generative models with latent variables, dedicated to model time series of high-dimensional data. DVAEs can be considered as extensions of the variational autoencoder (VAE) that include temporal dependencies between successive observed and/or latent vectors. Previous work has shown the interest of using DVAEs over the VAE for speech spectrograms modeling. Independently, the VAE has been successfully applied to speech enhancement in noise, in an unsupervised noise-agnostic set-up that requires neither noise samples nor noisy speech samples at training time, but only requires clean speech signals. In this paper, we extend these works to DVAE-based single-channel unsupervised speech enhancement, hence exploiting both speech signals unsupervised representation learning and dynamics modeling. We propose an unsupervised speech enhancement algorithm that combines a DVAE speech prior pre-trained on clean speech signals with a noise model based on nonnegative matrix factorization, and we derive a variational expectation-maximization (VEM) algorithm to perform speech enhancement.
The algorithm is presented with the most general DVAE formulation and is then applied with three specific DVAE models to illustrate the versatility of the framework. Experimental results show that the proposed DVAE-based approach outperforms its VAE-based counterpart, as well as several supervised and unsupervised noise-dependent baselines, especially when the noise type is unseen during training.
\end{abstract}


%
\IEEEpeerreviewmaketitle

\section{Introduction}

\IEEEPARstart{S}{peech} enhancement is a classical and fundamental problem in speech processing \cite{benesty2006speech, loizou2013speech}, which aims to recover the clean speech signal from a noisy recording. Classical signal-processing-based solutions include spectral subtraction  \cite{boll1979suppression} and Wiener filtering \cite{lim1979enhancement} (which use noise and  clean speech power spectral density estimates obtained from the noisy signal), and the short-time spectral amplitude estimator \cite{ephraim1984speech}. 

Recently, the advances in deep learning (DL) have opened new possibilities to tackle this task. The most widely studied approach to DL-based speech enhancement is that of a regression problem in the time-frequency (TF) domain where a deep neural network (DNN) is trained to map an input noisy speech signal into an output clean speech signal or into a denoising TF mask that is applied on the noisy signal (see a review in \cite{wang2018supervised}). We can refer to this general approach as a noisy-to-clean mapping (N2C). Recent works have considered N2C directly in the time domain instead of the TF domain \cite{fu2017raw} or leveraging generative adversarial networks (GANs) \cite{pascual2017segan,fu2019metricgan,fu2021metricgan+}.  

In the N2C approach, model training is typically done in a \textit{supervised} manner using a \textit{parallel} noisy-clean dataset, i.e., with noisy and clean versions of the same speech signal. Such parallel dataset must be prepared beforehand, the noisy version being obtained by summing the clean speech signal with noise. With a large amount of training data, DNNs can efficiently learn the N2C denoising mapping \cite{lu2013speech, wang2018supervised}. However, supervised methods, whether they work in the TF domain or in the time domain, tend to have difficulties for generalizing to noise types and acoustic conditions that were not seen during training. And it is difficult, not to say impossible, to generate a dataset that includes all possible types and levels of noise (e.g., urban noise vs. office noise) and all possible acoustic conditions (e.g., different recording equipments, varying mouth-to-microphone distance and orientation, different reverberation characteristics, etc.).

Recent works in DL-based speech enhancement have tried to relax the constraints regarding the degree of supervision to ease the design of datasets and/or improve the generalization capability of the models. In the present speech enhancement context, relaxing the degree of supervision means that we go from methods using carefully aligned parallel noisy-clean data to methods using \emph{non-parallel} noisy-clean data (i.e., the noisy examples are not the noisy version of the clean examples), which are easier to prepare, or noisy-only data, which are both easy to record and prepare, or clean-only data, which are not so easy to record, but which can lead to good generalization capabilities, as seen below. For example, the GANs employed in \cite{xiang2020parallel} and \cite{yu2021cyclegan} use non-parallel noisy and clean speech examples. In the present context, such methods that do not require a parallel noisy-clean dataset, are referred to as \emph{unsupervised}. They can be divided into two groups. 

\emph{Unsupervised noise-dependent} methods use noise examples or noisy speech examples only (they do not use clean speech signals). For example, a noisy-to-noisy (N2N) mapping approach, originally proposed for image denoising in \cite{lehtinen2018noise2noise}, was applied to speech enhancement in \cite{alamdari2021improving, kashyap2021speech}. In this approach, the DNN input is still a noisy signal but the output clean signal is replaced with another noisy version of the same clean signal. This is supported by theoretical considerations: If the noises in the noisy input and output are zero-mean and uncorrelated, and an infinite number of examples is provided to the DNN, the latter will learn to output an average denoised version of the signal. The motivation for adopting this approach for speech enhancement is that clean audio signals are difficult and expensive to record (in studio condition) compared to noisy speech signals. However, the required assumptions are not met for audio signals. Not only the different channels of multichannel recordings do not contain the exact same clean speech signal, but they also contain correlated noise \cite{van1988beamforming, benesty_microphone_2008, gannot2017consolidated, vincent2018audio}. Moreover, in their experiments, the authors of \cite{alamdari2021improving, kashyap2021speech} have to rely on simulated clean speech plus noise signals, which questions the interest of the N2N approach compared to the conventional N2C mapping. However, the N2N approach inspired the more realistic noisy-target training method (NyTT) \cite{fujimura2021noisy}, in which noisy speech and extra noise is used. In the training step, the NyTT input is noisy speech further corrupted by an additional noise and the network is trained to recover the noisy speech at the output. Then in the test step, the network is supposed to recover the clean speech signal from a noisy speech input. Although NyTT lacks theoretical support, it was shown to obtain good results in practice \cite{fujimura2021noisy}. In a different spirit, another approach not using clean speech signals is the MetricGAN-U method proposed by Fu~\etal~\cite{fu2021metricganU}, an unsupervised version of their previous model MetricGAN \cite{fu2019metricgan,fu2021metricgan+}. MetricGAN-U relies on the non-intrusive speech quality metric DNSMOS \cite{reddy2021dnsmos}, which does not require using the clean speech signal, in contrast to the intrusive PESQ metric \cite{rix2001perceptual} used in MetricGAN. One problem with the unsupervised noise-dependent methods in general is that they learn the noise characteristics and acoustic conditions, and thus may generalize poorly to unseen noise and acoustic conditions, just like supervised N2C methods.

Alternatively, \emph{unsupervised noise-agnostic} methods are based on a (deep) model of clean speech signals and do not learn the noise characteristics during training. Instead, the latter are estimated at test time on each speech sequence to denoise, hence conceptually letting the speech enhancement method the potential to adapt to any kind of noise. This setting was originally referred to as \textit{semi-supervised} in the audio source separation literature \cite{smaragdis2007supervised,mysore2011non,mohammadiha2013supervised}, because it exploits a dataset of isolated signals for one of the sources in the mixture. This dataset is thus labeled with the class of the sound source, e.g., clean speech for speech enhancement. In the present paper, we choose to call this setting \textit{unsupervised} because in the machine learning literature, semi-supervised  refers to methods that are trained from both labeled and unlabeled datasets (e.g., \cite{NIPS2014_d523773c}). In this context, a semi-supervised speech enhancement method would be trained from both a labeled dataset of noisy and clean speech signal pairs, and an unlabeled dataset containing only noisy or clean speech. While very interesting, this setting is not considered in this paper.

Examples of unsupervised noise-agnostic methods include that of Bando~\etal~\cite{bando2018statistical}, who proposed to use a variational autoencoder (VAE) \cite{kingma2013auto,rezende2014stochastic} to learn a prior distribution of the clean speech signals. At test time, the noise signal is modeled with Bayesian nonnegative matrix factorization (NMF) \cite{fevotte2009nonnegative} whose parameters, as well as the VAE latent variables, are estimated with a Markov chain Monte Carlo (MCMC) algorithm, including a sampling of the NMF parameters and the VAE latent variables. The same general approach was considered and extended in \cite{leglaive2018variance} within an expectation-maximization optimization framework, in \cite{leglaive2019speech} with an alpha-stable noise model, in \cite{pariente2019statistically} with efficient inference and learning algorithms, and in \cite{sekiguchi2018bayesian, leglaive2019semi, fontaine2019cauchy} for a multi-channel configuration. More recently, a guided VAE was proposed in \cite{carbajal2021guided}, where the clean speech prior is defined conditionally on a voice activity detection or an ideal binary mask. This guiding information is provided by a supervised classifier, separately trained on noisy speech signals. Other supervised extensions of the speech enhancement framework combining a VAE clean speech model and an NMF noise model include \cite{bando2020adaptive} and \cite{fang2021variational}.

Most of the above VAE-based unsupervised noise-agnostic speech enhancement methods focused on exploiting different distributions and algorithms. Very few works dealt with the inherent limitation of the VAE to handle sequential data correlated in time, as is the case of speech data. To the best of our knowledge, only two papers proposed generative approaches to speech enhancement based on VAE variants that can learn temporal dependencies: A recurrent VAE (RVAE) based on recurrent neural networks (RNNs) was proposed in \cite{leglaive2020recurrent} and stochastic temporal convolutional networks (TCNs) \cite{aksan2018stcn, lea2016temporal} were used in \cite{richter2020speech}, allowing the latent variables to have both hierarchical and temporal dependencies. Yet, a series of works have focused on developing extensions of the original VAE for time series (completely independently of the speech enhancement problem). The deep Kalman filter (DKF) \cite{krishnan2015deep, krishnan2017structured} is a DNN-based state-space model that combines a VAE with a non-linear first-order Markov model on the latent vectors. The variational RNN (VRNN) \cite{chung2015recurrent} and the stochastic RNN (SRNN) \cite{fraccaro2016sequential} are other temporal extensions of the VAE, with more complex temporal dependencies between the observed and latent data sequences, implemented with DNNs and RNNs. Actually, RVAE, DKF, VRNN, SRNN, and several other models \cite{fabius2014variational,bayer2014learning,yingzhen2018disentangled,fraccaro2017disentangled} can all be seen as particular instances of a general class of models called dynamical variational autoencoders (DVAEs), which have been recently reviewed in \cite{girin2020dynamical}. In the present paper, we propose an unsupervised noise-agnostic speech enhancement algorithm based on the modeling of the clean speech signal with a DVAE and the use of the variational inference methodology \cite{neal1998view,Wainwright2008,Bishop2006}. We present this algorithm in the general context of the DVAE class of models, and then we apply it on three specific DVAE models in our experiments: RVAE (extending our preliminary work in \cite{leglaive2020recurrent}), DKF, and SRNN. To our knowledge, the present paper is the first in-depth study on the use of DVAEs, as a general class of models, for unsupervised speech enhancement.

The rest of the paper is organized as follows. Section~\ref{sec:dvae} presents the technical background of DVAE models and their application to speech signals modeling. Section~\ref{sec:se} presents the proposed DVAE-based speech enhancement algorithm. Section~\ref{sec:experiments} presents a series of experiments conducted with the three example DVAE models and their comparison with several state-of-the-art supervised and unsupervised speech enhancement methods. This includes cross-dataset experiments that investigate the generalization capabilities of the methods to unseen types of noise. Section~\ref{sec:conclusion} concludes the paper.

\section{DVAE and Speech Modeling}
\label{sec:dvae}
In this section, we first review the standard VAE and its extensions to temporal models, which are referred to as DVAEs. Then, we briefly introduce speech modeling using DVAE models. In the end of this section, we describe the practical implementation of three typical DVAE models. 

\subsection{VAEs and DVAEs}
In a VAE \cite{kingma2013auto,rezende2014stochastic}, an observed variable $\mys$ of high dimension $F$ is assumed to be generated from an unobserved, or latent, random variable $\myz$ of low dimension $L \ll F$. Let $p_{\bs{\theta}}(\mys, \myz) = p_{\bs{\theta}_{\mys}}(\mys|\myz) p_{\bs{\theta}_{\myz}}(\myz)$ be the parametric generative model of their joint distribution, where $\bs{\theta}=\bs{\theta}_{\bs{s}}\cup\bs{\theta}_{\bs{z}}$ denotes the set of parameters. In general, the latent vector $\myz$ is assumed to be generated from a very simple prior distribution, typically the multivariate standard Gaussian distribution $p_{\bs{\theta}_{\myz}}(\myz) = \mcal{N}(\myz; \mbf{0},\mbf{I})$ (in that case, $\bs{\theta}_{\myz}=\emptyset$). The parameters of $p_{\bs{\theta}_{\mys}}(\mys|\myz)$ are provided by a complex nonlinear function of $\myz$, implemented with a deep neural network (DNN) (and $\bs{\theta}_{\mys}$ is the set of parameters of this DNN).

Given a dataset $\mbf{S}=\{\mys_n\}_{n=1}^N$ of $N$ i.i.d. samples of $\mys$, a probabilistic model is traditionally optimized by maximizing the log-marginal likelihood (also called evidence), $\log p_{\bs{\theta}}(\mbf{S}) = \textstyle \sum_{n=1}^N \log p_{\bs{\theta}} (\mys_n)$, over the parameter set $\bs{\theta}$. In the VAE case, the complexity of $p_{\bs{\theta}_{\mys}}(\mys | \myz)$ makes the marginalization over the latent variable, and thus the computation of $\log p_{\bs{\theta}}(\mbf{S})$, intractable, and the same for the  posterior distribution $p_{\bs{\theta}}(\myz | \mys)$. Therefore, instead of directly maximizing $ \log p_{\bs{\theta}}(\mbf{S})$, an inference model $q_{\bs{\phi}}(\myz | \mys) \approx p_{\bs{\theta}} (\myz | \mys)$ is introduced, which is also defined by a DNN (of parameters $\bs{\phi}$). Then, the following evidence lower bound (ELBO) is computed:
\begin{align}
        &\mcal{L}(\bs{\theta},\bs{\phi}; \mbf{S}) = \sum_{n=1}^N \mbb{E}_{q_{\bs{\phi}}(\myz_n | \mys_n)} \big[ \log p_{\bs{\theta}}(\mys_n,\myz_n) - \log q_{\bs{\phi}}(\myz_n | \mys_n) \big] \nonumber \\
        & \qquad \quad = \log p_{\bs{\theta}}(\mbf{S}) - \sum_{n=1}^N D_{\text{KL}} \big(q_{\bs{\phi}}(\myz_n | \mys_n) || p_{\bs{\theta}} (\myz_n | \mys_n)\big),
\label{eq:vae_elbo}
\end{align}
where $D_{\text{KL}}(.)$ denotes the Kullback-Leibler (KL) divergence, which is always non-negative. The generative model $p_{\bs{\theta}_{\mys}}(\mys | \myz)$ and the inference model $q_{\bs{\phi}} (\myz | \mys)$ are jointly trained by maximizing the ELBO with respect to $\bs{\theta}_{\mys}$ and $\bs{\phi}$, using stochastic gradient descent combined with sampling \cite{kingma2013auto,rezende2014stochastic}.

While the vanilla VAE assumes statistical independence among observation vectors, DVAEs can be seen as an extension of the VAE for modeling sequential data correlated in time \cite{girin2020dynamical}. A DVAE keeps the global encoder-decoder philosophy of the VAE, but considers a sequence of (high-dimensional) observed random vectors $\mys_{1:T} = \{\mys_t\}_{t=1}^T$ and a corresponding sequence of (low-dimensional) latent vectors $\myz_{1:T} = \{ \myz_t\}_{t=1}^T$. A DVAE is thus defined by the joint probability density function (pdf) of observed and latent sequences $p_{\bs{\theta}}(\mys_{1:T}, \mbf{z}_{1:T})$, which can be factorized using the chain rule:\footnote{Here and in all the following, we take the convention that $\mbf{s}_{1:0} = \mbf{z}_{1:0} = \emptyset$. For $t=1$ the first term of the product in \eqref{eq:dvae_generation_joint} and \eqref{eq:dvae_generation_factorized} is thus $p_{\bs{\theta}}(\mbf{s}_{1}, \mbf{z}_{1})$ and $p_{\bs{\theta}}(\mbf{s}_{1} | \mbf{z}_{1})p_{\bs{\theta}}(\mbf{z}_{1})$, respectively.}
\begin{align}
p_{\bs{\theta}}(\mbf{s}_{1:T}, \mbf{z}_{1:T}) &= \prod_{t=1}^{T} p_{\bs{\theta}}(\mbf{s}_{t}, \mbf{z}_{t} | \mbf{s}_{1:t-1},\mbf{z}_{1:t-1})
\label{eq:dvae_generation_joint} \\
&= \prod_{t=1}^{T} p_{\bs{\theta}_{\mys}}(\mbf{s}_{t} | \mbf{s}_{1:t-1},\mbf{z}_{1:t}) p_{\bs{\theta}_{\myz}}(\mbf{z}_{t} | \mbf{s}_{1:t-1},\mbf{z}_{1:t-1}).
\label{eq:dvae_generation_factorized}
\end{align}
As for the VAE, the exact posterior distribution $p_{\bs{\theta}}(\mbf{z}_{1:T} | \mbf{x}_{1:T})$ is not analytically tractable. Consequently, an approximate posterior distribution $q_{\bs{\phi}}(\mbf{z}_{1:T} | \mbf{x}_{1:T})$ is introduced and it can be factorized using the chain rule as:
\begin{align}
q_{\bs{\phi}}(\mbf{z}_{1:T} | \mbf{s}_{1:T}) &= \prod_{t=1}^{T} q_{\bs{\phi}}(\mbf{z}_{t} | \mbf{z}_{1:t-1},\mbf{s}_{1:T}).
\label{eq:dvae_inference}
\end{align}
Chaining the inference and generation, the training of DVAEs is done by maximizing the ELBO on a set of training vector sequences, the ELBO being here defined by (for a single observed and latent data sequence):
\begin{equation}
    \begin{split}
        \mcal{L}(\bs{\theta},\bs{\phi}; \mys_{1:T}) = \mbb{E}_{q_{\bs{\phi}}(\mbf{z}_{1:T} | \mbf{s}_{1:T})} &\left[ \ln p_{\bs{\theta}}(\mbf{s}_{1:T},\mathbf{z}_{1:T} ) \right. \\
        & - \left.  \ln q_{\bs{\phi}}(\mathbf{z}_{1:T} | \mbf{s}_{1:T}) \right].
    \end{split}
\label{eq:dvae_elbo}
\end{equation}

When writing a joint distribution as a product of conditional distributions using the chain rule, a specific ordering of the variables has to be chosen. Among different possibilities, we chose a \emph{causal} ordering to write the factorization in (\ref{eq:dvae_generation_joint}) and (\ref{eq:dvae_generation_factorized}): The generation of $\mys_t$ and $\myz_t$ uses their past values $\mys_{1:t-1}$ and $\myz_{1:t-1}$ (plus $\myz_{t}$ for generating $\myx_{t}$). In the DVAE literature, almost all models are causal \cite{girin2020dynamical}. Each of them can be seen as a special case of the general expression (\ref{eq:dvae_generation_factorized}) where the dependencies in $p_{\bs{\theta}}(\mbf{s}_{t} | \mbf{s}_{1:t-1},\mbf{z}_{1:t})$ and $p_{\bs{\theta}}(\mbf{z}_{t} | \mbf{s}_{1:t-1},\mbf{z}_{1:t-1})$ are simplified, which may also affect the choice of the inference model $q_{\bs{\phi}}(\mbf{z}_{1:T} | \mbf{s}_{1:T})$ in \eqref{eq:dvae_inference}. In addition, a given DVAE model can have different implementations with various types of DNNs, see \cite{girin2020dynamical} for an extensive discussion on this topic.

\subsection{Speech modeling using DVAEs}
\label{subsec:speechModelingUsingDVAEs}

\begin{figure*}[thbp]
    \centering
    \includegraphics[width=1\linewidth]{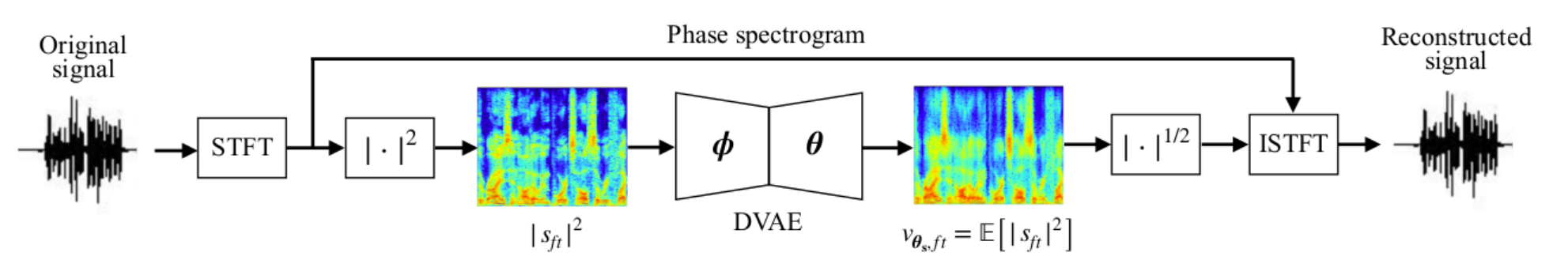}
    \caption{Speech analysis-resynthesis with a DVAE model in the STFT domain. The speech power spectrogram is used as input to the DVAE and the output is the variance of a complex Gaussian model. The audio waveform is reconstructed by inverse STFT using the phase of the original signal.}
    \label{fig:resynthesis}
\end{figure*}

The VAE and the DVAE class of models have been used to model different kinds of data. In this subsection, we discuss the use of DVAEs for modeling speech signals in the short-term Fourier transform (STFT) domain. Fig.~\ref{fig:resynthesis} illustrates this process. Let $\mbf{s}_{1:T} = \{\mbf{s}_{t} \in \mathbb{C}^F\}_{t=1}^T$ denote an $F \times T$ sequence of complex-valued STFT frames, where $t$ is the time-frame index. Each vector $\mbf{s}_{t} = \{s_{ft} \in \mathbb{C} \}_{f=1}^{F}$ represents the speech short-term spectrum at time index $t$, and $f$ is the frequency bin. As indicated above, $\mbf{s}_{1:T}$ is associated with an $L \times T$ sequence of latent variables $\mbf{z}_{1:T} = \{\mbf{z}_{t} \in \mathbb{R}^L\}_{t=1}^T$, with $L \ll F$. 

In speech and audio processing, the Fourier coefficients in $\mbf{s}_{t} \in \mathbb{C}^{F}$ are usually assumed to be independent and distributed according to a  complex Gaussian circularly symmetric distribution \cite{neeser1993proper} (denoted below by $\mcal{N}_c$), whose variance vary over time and frequency \cite{ephraim1984speech, vincent2011probabilistic} (the circularly symmetric assumption means that the phase follows a uniform distribution in $[0, 2\pi)$). Thus, for all time frames $t \in \{1,\dots,T\}$, the DVAE generative model of speech signal is defined as follows:
\begin{align}
    & p_{\bs{\theta}_{\mys}}(\mbf{s}_{t} | \mbf{s}_{1:t-1},\mbf{z}_{1:t}) = \mcal{N}_c\left(\mbf{s}_{t}; \mbf{0},  \bs{\Sigma}_{\bs{\theta}_{\mbf{s}},t}\right),
    \label{eq:speech_gen_s} \\
    & p_{\bs{\theta}_{\myz}}(\mbf{z}_{t} | \mbf{s}_{1:t-1},\mbf{z}_{1:t-1}) = \mcal{N}\left(\mbf{z}_{t};\bs{\mu}_{\bs{\theta}_{\mbf{z}},t}, \bs{\Sigma}_{\bs{\theta}_{\mbf{z}},t}\right),
    \label{eq:speech_gen_z}
\end{align}
where the diagonal covariance matrix $\bs{\Sigma}_{\bs{\theta}_{\mbf{s}},t} = \diag\{\bs{v}_{\bs{\theta}_{\mbf{s}},t}\}$ is provided by a DNN that takes as input the conditioning variables in \eqref{eq:speech_gen_s}, namely $(\mbf{s}_{1:t-1}, \mbf{z}_{1:t})$. Similarly, $\bs{\mu}_{\bs{\theta}_{\mbf{z}},t}$ and $\bs{\Sigma}_{\bs{\theta}_{\mbf{z}},t} = \diag\{\bs{v}_{\bs{\theta}_{\mbf{z}},t}\}$ are provided by a DNN that takes as input the conditioning variables in \eqref{eq:speech_gen_z}, namely $(\mbf{s}_{1:t-1}, \mbf{z}_{1:t-1})$.\footnote{It is important to note that, in DVAEs, a parameter of a distribution is always a function of the variables that come after the conditioning bar. In the rest of the paper, we will generally omit to rewrite these variables in the right-hand-side of the probabilistic modeling equations, for concision, but we may punctually make these dependencies explicit when it eases the understanding.} We denote by $\bs{\theta} = \bs{\theta}_{\bs{s}}\cup\bs{\theta}_{\bs{z}}$ the parameters of the DNNs involved in \eqref{eq:speech_gen_s} and \eqref{eq:speech_gen_z}. 

As for the inference model, it is given by:
\begin{equation}
    q_{\bs{\phi}} (\myz_t | \mbf{z}_{1:t-1}, \mys_{1:T}) = \mcal{N}(\mbf{z}_{t};\bs{\mu}_{\bs{\phi},t}, \bs{\Sigma}_{\bs{\phi},t}),
    \label{eq:speech_inf}
\end{equation}
where $\bs{\mu}_{\bs{\phi},t}$ and $\bs{\Sigma}_{\bs{\phi},t} = \diag\{\bs{v}_{\bs{\phi},t}\}$ are provided by a DNN taking $(\mbf{z}_{1:t-1}$, $\mys_{1:T})$ as input and whose parameters are denoted by $\bs{\phi}$.

Even if the complex-valued vector sequence $\mbf{s}_{1:t-1}$ or $\mbf{s}_{1:T}$ is used as a conditioning variable in \eqref{eq:speech_gen_s}-\eqref{eq:speech_inf}, in practice we use the modulus-squared values of these variables at the encoder and decoder input. In other words, the DVAE encoder and decoder distribution parameters are computed using sequences of vectors with entries equal to $|s_{ft}|^2$, as illustrated in Fig.~\ref{fig:resynthesis}. Note that the modulus-squared of data is homogeneous with the decoder output (the variance vector $\bs{v}_{\bs{\theta}_{\mbf{s}},t}$).

\begin{figure*}[thbp]
    \centering
    \includegraphics[width=0.9\linewidth]{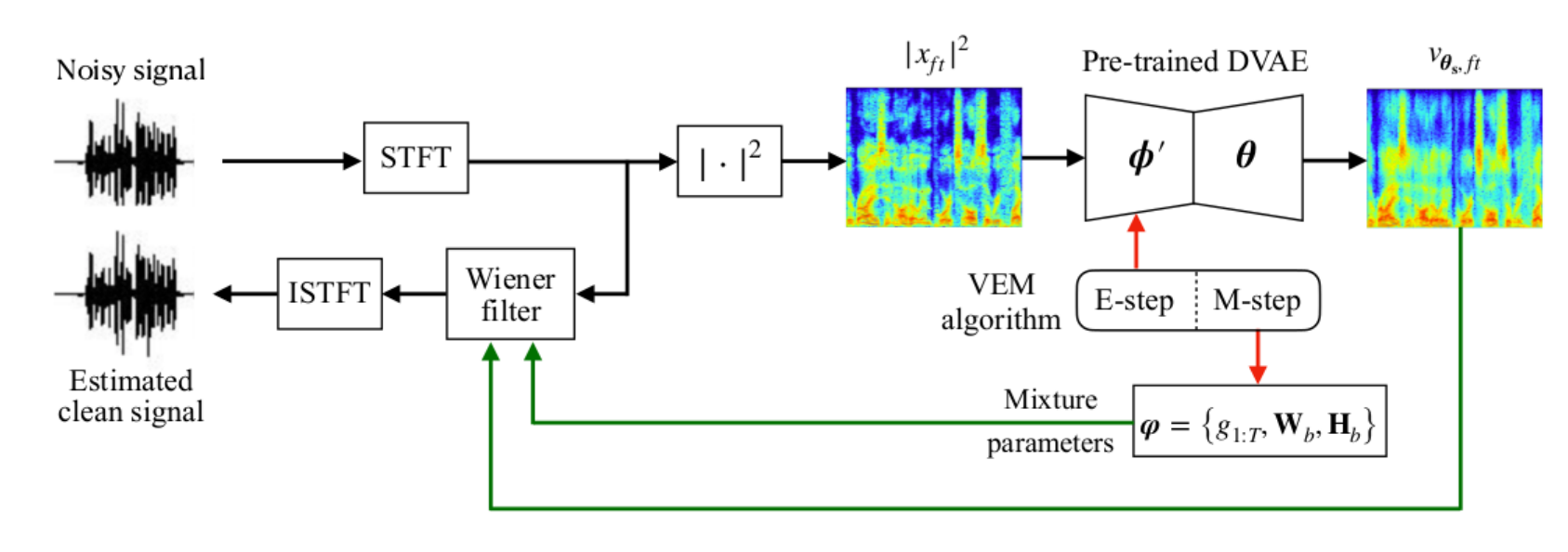}
    \caption{Overview of the proposed speech enhancement method. The pre-trained DVAE is used within a VEM algorithm for speech enhancement. The DVAE encoder is fine-tuned during the E-step and the mixture parameters are estimated in the M-step (red arrows), see Section~\ref{subsec:c} and Algorithm~1. The clean speech signal is estimated by filtering the noisy signal with a Wiener filter combining the DVAE output parameters and the estimated mixture parameters (green arrows).}
    \label{fig:enhancement}
\end{figure*}

Given the generative model (\ref{eq:speech_gen_s}), (\ref{eq:speech_gen_z}) and the inference model (\ref{eq:speech_inf}), we can develop the ELBO in \eqref{eq:dvae_elbo} as follows (for one data sequence):
\begin{align}
&\mcal{L}(\bs{\theta},\bs{\phi}; \mys_{1:T}) \nonumber \overset{c}{=} - \sum_{f,t=1}^{F,T} \mbb{E}_{q_{\bs{\phi}}(\mbf{z}_{1:t} | \mbf{s}_{1:T} )} \left[ d_{\text{IS}}(|s_{ft}|^2, v_{\bs{\theta}_{\mbf{s}},ft}) \right]\nonumber  \\
& \qquad +\frac{1}{2} \sum_{l,t=1}^{L,T} \Bigg[\ln \frac{v_{\bs{\phi},lt}}{v_{\bs{\theta}_{\mbf{z}},lt}} - \frac{v_{\bs{\phi},lt} + ({\mu}_{\bs{\phi},lt} - {\mu}_{\bs{\theta}_{\mbf{z}},lt})^2}{v_{\bs{\theta}_{\mbf{z}},lt}}\Bigg],
\label{eq:speech_elbo_factorized}
\end{align}
where $\overset{c}{=}$ denotes equality up to an additive constant w.r.t. $\bs{\theta}$ and $\bs{\phi}$, $d_{\text{IS}}(q,p) = q/p - \ln(q/p) -1$ is the Itakura-Saito (IS) divergence \cite{fevotte2009nonnegative}, $v_{\bs{\theta}_{\mys},ft} \in \mbb{R}_+$ is the $f$-th entry of $\bs{v}_{\bs{\theta}_{\mys},t}$, and $\{{\mu}_{\bs{\phi},lt} \in \mbb{R}, {\mu}_{\bs{\theta}_{\myz},lt} \in \mbb{R}, v_{\bs{\phi},lt} \in \mbb{R}_+, v_{\bs{\theta}_{\myz},lt} \in \mbb{R}_+\}$ are the $l$-th entry of $\{\bs{\mu}_{\bs{\phi},t}, \bs{\mu}_{\bs{\theta}_{\myz},t}, \bs{v}_{\bs{\phi},t}, \bs{v}_{\bs{\theta}_{\myz},t}\}$, respectively. 

\subsection{Three representative DVAE models}
\label{subsec:d}

In this subsection, we briefly present three DVAE models, namely DKF, RVAE, and SRNN, which are representative of the versatility of the DVAE class (see the review in \cite{girin2020dynamical}) and which we used in practice in our speech enhancement method, with results reported in Section~\ref{sec:experiments}.

\subsubsection{{DKF}} As briefly stated in the introduction, DKF  \cite{krishnan2015deep,krishnan2017structured} is the simplest DVAE model, following the structure of a basic state-space model with a first-order Markov model on the latent vectors, i.e., $\myz_t$ is generated from $\myz_{t-1}$, and an instantaneous observation model, i.e., $\mys_t$ is generated from $\myz_t$:
\begin{align}
    p_{\mytheta}(\mys_{1:T}, \myz_{1:T}) &= \prod_{t=1}^{T} p_{\mytheta_{\mys}} (\mys_t | \myz_t) p_{\mytheta_{\myz}}(\myz_t | \myz_{t-1}).
    \label{eq:dkf_gen}
\end{align}
The inference model is defined by:
\begin{align}
    q_{\myphi} (\myz_{1:T} | \mys_{1:T}) &= \prod_{t=1}^{T} q_{\myphi} (\myz_t | \myz_{t-1}, \mys_{t:T}).  \label{eq:dkf_inf}
\end{align}

\subsubsection{{RVAE}} As opposed to DKF, RVAE \cite{leglaive2020recurrent} does not consider any dynamical model for the latent vectors, which are assumed i.i.d., with $p(\myz_t) = \mathcal{N}(\mbf{z}_t; \mbf{0}, \mbf{I)}$. This prior therefore lacks parameters and does not involve a neural network, as in standard VAEs. However, the observation model in RVAE is more complex than in DKF, as $\mys_t$ is generated from the current and previous latent variables $\mbf{z}_{1:t}$. Formally, the RVAE generative and inference models are defined by:
\begin{align}
     p_{\mytheta}(\mys_{1:T}, \myz_{1:T}) &= \prod_{t=1}^{T} p_{\mytheta_{\mys}} (\mys_t | \myz_{1:t}) p (\myz_t), \label{eq:rvae_causal_gen}\\
     q_{\myphi} (\myz_{1:T} | \mys_{1:T}) &= \prod_{t=1}^{T} q_{\myphi} (\myz_t | \myz_{1:t-1}, \mys_{t:T}). \label{eq:rvae_causal_inf}
\end{align}
RVAE is the only model that was presented in the literature with both a causal and non-causal form \cite{leglaive2020recurrent,girin2020dynamical}. The non-causal form is obtained by simply replacing $\mbf{z}_{1:t}$ with $\mbf{z}_{1:T}$ in \eqref{eq:rvae_causal_gen} and by replacing $\mys_{t:T}$ with $\mys_{1:T}$ in \eqref{eq:rvae_causal_inf}.

\subsubsection{{SRNN}} In SRNN \cite{fraccaro2016sequential}, the latent vector $\myz_t$ is generated not only from $\myz_{t-1}$ (as in DKF), but also from $\mys_{1:t-1}$. And $\mys_t$ is generated not only from $\myz_t$ (as in DKF) but also from $\mys_{1:t-1}$, so that SRNN actually corresponds to an autoregressive model. The generative and inference models are defined by:
\begin{align}
    p_{\mytheta} (\mys_{1:T}, \myz_{1:T}) &= \prod_{t=1}^{T} p_{\mytheta_{\mys}} (\mys_t | \myz_t, \mys_{1:t-1}) p_{\mytheta_{\myz}} (\myz_t | \myz_{t-1}, \mys_{1:t-1}), \label{eq:srnn_gen}\\
    q_{\myphi} (\myz_{1:T} | \mys_{1:T}) &= \prod_{t=1}^T q_{\myphi} (\myz_t | \myz_{t-1}, \mys_{1:T}). \label{eq:srnn_inf}
\end{align}
\section{DVAE for Speech Enhancement}
\label{sec:se}
This section presents the proposed unsupervised noise-agnostic DVAE-based speech enhancement algorithm, where the clean speech signal is modeled with a DVAE and the noise is modeled with nonnegative matrix factorization (NMF) \cite{fevotte2009nonnegative}. It is an extended version of the algorithm proposed in \cite{leglaive2020recurrent} for the RVAE model, with a more general formulation applicable to any other DVAE model. The proposed method is illustrated in Fig. \ref{fig:enhancement}. We assume that the DVAE-based clean speech generative and inference models defined in \eqref{eq:speech_gen_s}-\eqref{eq:speech_gen_z} and \eqref{eq:speech_inf}, respectively, have been learned, i.e., the DNN parameters $\mytheta$ and $\bs{\phi}$ have been estimated from a dataset of clean speech signals during an independent training stage (see Section~\ref{subsec:speechModelingUsingDVAEs}). The objective of speech enhancement is to use this pre-trained DVAE model to estimate the clean speech signal when only the noisy mixture is observed. This is done with a variational expectation-maximization (VEM) algorithm \cite{neal1998view,Wainwright2008,Bishop2006}. We recall that this method is unsupervised, since no pair of clean and noisy speech examples are used. Moreover it is noise-agnostic since it does not make any assumption on the noise type, except that it can be modeled with an NMF model, and the noise model NMF parameters are estimated independently for each noisy sequence to process.   

In the rest of this section, we first introduce the noise and mixture models, then we develop the general strategy to estimate the clean speech signal modeled by a DVAE when only the mixture signal is available, and finally we present the VEM algorithm used to estimate the remaining unknown model parameters. Throughout this section, $\mys_{1:T} = \{\mys_t \in \mbb{C}^F \}_{t=1}^T$, $\mbf{b}_{1:T} = \{\mbf{b}_t \in \mbb{C}^F \}_{t=1}^T$, and $\myx_{1:T} = \{\myx_t \in \mbb{C}^F \}_{t=1}^T$ respectively denote the STFT of the clean speech signal, the noise signal, and the noisy speech signal.

\subsection{Noise and mixture models}
\label{subsec:a}
As in \cite{leglaive2018variance,leglaive2020recurrent}, we consider a Gaussian noise model with NMF parameterization of the variance \cite{fevotte2009nonnegative}. Independently for all time frames $t \in \{1,...,T\}$, we define:
\begin{equation}
p(\mbf{b}_t) = \mathcal{N}_c(\mbf{b}_t; \mbf{0}, \bs{\Sigma}_{\mbf{b},t}),
\label{eq:nmf_noise}
\end{equation}
where $\bs{\Sigma}_{\mbf{b},t} = \diag\{ (\mbf{W}_b \mbf{H}_b)_{:,t} \}$ with $\mbf{W}_b \in \mbb{R}_+^{F \times K}$ and $\mbf{H}_b \in \mbb{R}_+^{K \times T}$. The rank of the factorization $K$ is usually chosen such that $K(F + T) \ll FT$.

We consider that the noisy speech is a mixture of the noise defined in (\ref{eq:nmf_noise}) and the clean speech defined in (\ref{eq:speech_gen_s}) and (\ref{eq:speech_gen_z}):
\begin{equation}
    \mbf{x}_t = \sqrt{g_t}\mbf{s}_t + \mbf{b}_t,
    \label{eq:mixture0}
\end{equation}
where $g_t \in \mbb{R}_+$ is a frame-dependent frequency-independent gain parameter scaling the speech signal level at each time frame. This parameter enables to take into account the potentially different loudness between the clean speech training examples used to learn the DVAE model and the speech signal in the test noisy sequence we have to denoise \cite{leglaive2018variance}.

From \eqref{eq:speech_gen_s}, \eqref{eq:nmf_noise} and \eqref{eq:mixture0}, and by assuming the independence of the speech and noise signals, we have for all $t \in \{1,...,T\}$:
\begin{equation}
p_{\mytheta_\mbf{x}} (\mbf{x}_t \mid \mbf{s}_{1:t-1}, \mbf{z}_{1:t}) = \mathcal{N}_c\left(\mbf{x}_t; \mbf{0}, \bs{\Sigma}_{\mytheta_{\mbf{x}},t} \right),
\label{eq:mixture1}
\end{equation}
where $\bs{\Sigma}_{\mytheta_{\mbf{x}},t} = \diag\{ g_t \bs{v}_{\mytheta_{\mys},t} + (\mbf{W}_b \mbf{H}_b)_{:,t} \}$ and $\mytheta_{\myx}$ is the union of the speech generative model parameters $\mytheta_{\mys}$ and the mixture model parameters  $\myvarphi =\{\bs{g} = [g_1, \dots, g_T]^T,\mbf{W}_b,\mbf{H}_b\}$. As already mentioned in Section~\ref{subsec:speechModelingUsingDVAEs},  $\bs{v}_{\mytheta_{\mys},t}$ is actually a function of $(\mbf{s}_{1:t-1}, \mbf{z}_{1:t})$. 
Note that it is clear from \eqref{eq:nmf_noise} and \eqref{eq:mixture0} that given the clean speech frame $\mbf{s}_t$, the noisy speech frame $\mbf{x}_t$ is characterized by:
\begin{equation}
    p_{\myvarphi} (\mbf{x}_t \mid \mbf{s}_t) = \mathcal{N}_c\left(\mbf{x}_t; \sqrt{g_t}\mbf{s}_t, \bs{\Sigma}_{\mbf{b},t} \right).
\label{eq:mixture2}
\end{equation}

\subsection{Speech reconstruction}
\label{subsec:b}

Now we consider the problem of reconstructing the clean speech signal from the observed mixture signal, which consists in computing the following posterior mean vector:
\begin{equation}
    \hat{\mbf{s}}_{t} = \mbb{E}_{p_{\mytheta}(\mbf{s}_{t} | \mbf{x}_{1:T})} [\mbf{s}_{t}].
    \label{eq:speech_recon}
\end{equation}
However, we cannot write the posterior $p_{\mytheta}(\mbf{s}_{t} | \mbf{x}_{1:T})$ analytically, which makes the above expectation intractable. However, leveraging the speech model defined previously, we can approximate it by introducing random variables that are then marginalized.

\subsubsection{{Introducing the past and current latent variables}}

We start from marginalizing with respect to $\mbf{z}_{1:t}$:
\begin{align}
    p_{\mytheta}(\mbf{s}_{t} | \mbf{x}_{1:T}) &= \int p_{\mytheta}(\mbf{s}_t | \mbf{z}_{1:t}, \mbf{x}_{1:T}) p_{\mytheta}(\mbf{z}_{1:t} | \mbf{x}_{1:T}) d \mbf{z}_{1:t} \nonumber\\
    &= \mbb{E} _{p_{\mytheta}(\mbf{z}_{1:t} | \mbf{x}_{1:T} )} [p_{\mytheta}(\mbf{s}_t | \mbf{z}_{1:t}, \mbf{x}_{1:T})].
    \label{eq:recon_MargiZ}
\end{align}
Using \eqref{eq:recon_MargiZ} to rewrite \eqref{eq:speech_recon}, the estimate of the clean speech signal at time $t$ is given by:
\begin{equation}
    \hat{\mbf{s}}_{t} = \mbb{E}_{p_{\mytheta}(\mbf{z}_{1:t} | \mbf{x}_{1:T} )} \left[ \mbb{E}_{p_{\mytheta}(\mbf{s}_t | \mbf{z}_{1:t}, \mbf{x}_{1:T})} [\mbf{s}_{t}] \right].
    \label{eq:speech_recon2}
\end{equation}
Let us now focus on the inner expectation, taken with respect to $p_{\mytheta}(\mbf{s}_t | \mbf{z}_{1:t}, \mbf{x}_{1:T})$. We will come back later on the outer expectation taken with respect to $p_{\mytheta}(\mbf{z}_{1:t} | \mbf{x}_{1:T} )$. Using Bayes rule, we have:
\begin{align}
    p_{\mytheta}(\mbf{s}_t | \mbf{z}_{1:t}, \mbf{x}_{1:T}) & = \frac{p_{\mytheta}(\mbf{x}_{1:T} | \mbf{s}_t, \mbf{z}_{1:t}) p_{\mytheta}(\mbf{s}_t | \mbf{z}_{1:t}) p_{\mytheta}(\mbf{z}_{1:t}) }{p_{\mytheta}( \mbf{z}_{1:t}, \mbf{x}_{1:T}) } \\
    & \propto p_{\mytheta}(\mbf{x}_{1:T} | \mbf{s}_t, \mbf{z}_{1:t}) p_{\mytheta}(\mbf{s}_t | \mbf{z}_{1:t})     \label{eq:se_method1_bayesian-a}
\\
    & \approx p_{\mytheta}(\mbf{x}_{t} | \mbf{s}_t) p_{\mytheta}(\mbf{s}_t | \mbf{z}_{1:t}).
    \label{eq:se_method1_bayesian}
\end{align}
The exact computation of $p_{\mytheta}(\mbf{x}_{1:T} | \mbf{s}_t, \mbf{z}_{1:t})$ requires the marginalisation of $p_{\mytheta}(\mbf{x}_{1:T},\mbf{s}_{1:t-1,t+1:T}, \mbf{z}_{t+1:T} | \mbf{s}_t, \mbf{z}_{1:t})$ w.r.t.\ the undesired variables. This would require not only marginalising from future latent codes, but also from past and future clean speech, which is clearly not feasible. Instead, we approximate  \eqref{eq:se_method1_bayesian-a} with \eqref{eq:se_method1_bayesian} by considering only the signal mixture model $p_{\mytheta}(\mbf{x}_{t} | \mbf{s}_t)$, as defined in \eqref{eq:mixture2}.

\subsubsection{{Introducing the past speech vectors}}
\label{subsubsec:pastSpeechVectMarg}
Then, it comes to estimating $p_{\mytheta}(\mbf{s}_t | \mbf{z}_{1:t})$ in \eqref{eq:se_method1_bayesian}. To do so, we introduce and then marginalize the past speech vectors $\mbf{s}_{1:t-1}$:
\begin{align}
    p_{\mytheta}& (\mbf{s}_t | \mbf{z}_{1:t}) = \int p_{\mytheta}(\mbf{s}_t | \mbf{s}_{1:t-1}, \mbf{z}_{1:t}) p_{\mytheta} ( \mbf{s}_{1:t-1} | \mbf{z}_{1:t}) d \mbf{s}_{1:t-1} \nonumber\\
    &= \int p_{\mytheta}(\mbf{s}_t | \mbf{s}_{1:t-1}, \mbf{z}_{1:t}) \left[\prod_{\tau=1}^{t-1} p_{\mytheta}(\mbf{s}_{\tau} | \mbf{s}_{1:\tau-1}, \mbf{z}_{1:\tau}) \right] d \mbf{s}_{1:t-1} 
    \nonumber\\
    &= \mbb{E} _{\prod_{\tau=1}^{t-1} p_{\mytheta}(\mbf{s}_{\tau} | \mbf{s}_{1:\tau-1}, \mbf{z}_{1:\tau})} \left[\,p_{\mytheta}(\mbf{s}_t | \mbf{s}_{1:t-1}, \mbf{z}_{1:t})\, \right],
    \label{eq:se_sample_s}
\end{align}
where in the second line we used the fact that $\mys_{\tau}$ is conditionally independent of $\myz_{\tau+1 : t}$.

When computing \eqref{eq:se_sample_s}, we are facing two problems: First, the expectation is intractable; and second, in a speech enhancement framework, we do not have access to the past ground-truth clean speech vectors  $\mbf{s}_{1:t-1}$ (as opposed to the DVAE training procedure which is done using sequences of clean speech signals). Therefore, we approximate $p_{\mytheta}(\mbf{s}_t | \mbf{z}_{1:t})$ as follows:
\begin{equation}
    \begin{split}
    p_{\mytheta}(\mbf{s}_t | \mbf{z}_{1:t}) &\approx p_{\mytheta}(\mbf{s}_t | \tilde{\mbf{s}}_{1:t-1}, \mbf{z}_{1:t}) \\
    &= \mathcal{N}_c\big(\mbf{s}_t; \mbf{0}, \bs{\Sigma}_{\mytheta_{\mbf{s}},t}(\tilde{\mbf{s}}_{1:t-1}, \mbf{z}_{1:t})\big),
    \end{split}
\label{eq:se_method1_approx_s}
\end{equation}
where $\bs{\Sigma}_{\mytheta_{\mbf{s}},t}(\tilde{\mbf{s}}_{1:t-1}, \mbf{z}_{1:t}) = \diag\{\bs{v}_{\mytheta_{\mbf{s}},t}(\tilde{\mbf{s}}_{1:t-1}, \mbf{z}_{1:t})\}$ and $\tilde{\mbf{s}}_{t}$ is computed recursively as $\tilde{\mbf{s}}_{t} = \bs{v}_{\mytheta_{\mbf{s}},t}(\tilde{\mbf{s}}_{1:t-1}, \mbf{z}_{1:t})$.\footnote{Here we explicitly write the dependency of the covariance matrix on $\tilde{\mbf{s}}_{1:t-1}$ and $\mbf{z}_{1:t}$, to make the use of the DVAE model clear. In the following we will omit it again for concision of presentation.} In practice, the decoder output at time frame $t-1$ is re-injected at the decoder input at the next time frame $t$.
This part of the process is necessary only for SRNN, and more generally for any autoregressive DVAE. For non-autoregressive DVAE models, such as RVAE and DKF, $\bs{\Sigma}_{\mytheta_{\mbf{s}},t}$ is only computed from the sequence of latent vectors.

\subsubsection{{Computing the conditional posterior}}
Substituting (\ref{eq:mixture2}) and (\ref{eq:se_method1_approx_s}) into (\ref{eq:se_method1_bayesian}), we have:
\begin{align}
    p_{\mytheta}(\mbf{s}_t | \mbf{z}_{1:t}, \mbf{x}_{1:T}) &\approx \mathcal{N}_c \left( \mbf{x}_t; \sqrt{g_t} \mbf{s}_t, \bs{\Sigma}_{\mbf{b},t}\right) \mathcal{N}_c \left( \mbf{s}_t; \mbf{0}, \bs{\Sigma}_{\mytheta_{\mbf{s}},t}\right) \nonumber \\
    &= \mathcal{N}_c \left( \mbf{s}_t; \mbf{m}_{\mbf{s},t},  \bs{\Sigma}_{\mbf{s},t} \right),
    \label{eq:cond_post_s}
\end{align}
where
\begin{align}
    & \mbf{m}_{\mbf{s},t} =  \sqrt{g_t} \bs{\Sigma}_{\mytheta_{\mbf{s}},t} \left( g_t \bs{\Sigma}_{\mytheta_{\mbf{s}},t}  + \bs{\Sigma}_{\mbf{b},t} \right)^{-1} \mbf{x}_t, \label{eq:wiener_filter} \\
    & \bs{\Sigma}_{\mbf{s},t} = \bs{\Sigma}_{\mytheta_{\mbf{s}},t} \bs{\Sigma}_{\mbf{b},t} ( g_t \bs{\Sigma}_{\mytheta_{\mbf{s}},t} + \bs{\Sigma}_{\mbf{b},t})^{-1}.
\end{align}
Finally, from \eqref{eq:speech_recon2}, \eqref{eq:cond_post_s} and \eqref{eq:wiener_filter}, the estimate of the clean speech signal is given by:
\begin{align}
    \hat{\mbf{s}}_{t} & \approx \mbb{E}_{p_{\mytheta}(\mbf{z}_{1:t} | \mbf{x}_{1:T} )} \left[ \sqrt{g_t} \bs{\Sigma}_{\mytheta_{\mbf{s}},t} \left( g_t \bs{\Sigma}_{\mytheta_{\mbf{s}},t}  + \bs{\Sigma}_{\mbf{b},t} \right)^{-1} \right]\mbf{x}_t,
    \label{eq:se_approxS}
\end{align}
where we recall that $\bs{\Sigma}_{\mytheta_{\mbf{s}},t}$ is actually a function of $(\tilde{\mbf{s}}_{1:t-1}, \mbf{z}_{1:t})$. This speech signal estimate can be seen as the output of a ``probabilistic'' Wiener filter, i.e., a Wiener filter averaged over all possible realizations of the latent variables according to their posterior distribution $p_{\mytheta}(\mbf{z}_{1:t} | \mbf{x}_{1:T} )$.

The expectation in \eqref{eq:se_approxS} is intractable, but similarly as before we can approximate it by
\begin{equation}
    \hat{\mbf{s}}_{t} \approx \sqrt{g_t} \bs{\Sigma}_{\mytheta_{\mbf{s}},t} \left( g_t \bs{\Sigma}_{\mytheta_{\mbf{s}},t}  + \bs{\Sigma}_{\mbf{b},t} \right)^{-1} \mbf{x}_t,
    \label{eq:se_complete}
\end{equation}
where $\bs{\Sigma}_{\mytheta_{\mbf{s}},t} = \bs{\Sigma}_{\mytheta_{\mbf{s}},t}(\tilde{\mbf{s}}_{1:t-1},\tilde{\mbf{z}}_{1:t})$ and $\tilde{\mbf{z}}_{1:t}$ is sampled from $p_{\mytheta}(\mbf{z}_{1:t} | \mbf{x}_{1:T} )$. This posterior distribution is also intractable. We thus propose to use instead a variational approximation $q_{\myphi'} (\myz_{1:T} | \myx_{1:T})$ whose parameters $\myphi'$ need to be jointly estimated together with the noisy mixture model parameters $\myvarphi$. In the next section, we propose a VEM algorithm to do that. This generalizes the algorithm developed for RVAE in \cite{leglaive2020recurrent} to the whole class of DVAE models.

\subsection{VEM algorithm for model parameters estimation}
\label{subsec:c}

Now that we have an expression for the clean speech signal estimate, what remains to be estimated is the set of mixture model parameters $\myvarphi$ (the NMF noise model parameters and the gains) and the parameters $\myphi'$ of the variational distribution $q_{\myphi'} (\myz_{1:T} | \myx_{1:T})$. Using a VEM algorithm, we will maximize the following ELBO defined for the noisy speech observations $\myx_{1:T}$:
\begin{align}
\mcal{L}(\myphi',\myvarphi; \myx_{1:T}) = \mbb{E}_{q_{\myphi'}(\myz_{1:T}|\myx_{1:T})} [& \ln p_{\myvarphi}(\myx_{1:T}, \myz_{1:T}) \nonumber \\
&  - \ln q_{\myphi'}(\myz_{1:T} | \myx_{1:T})].
\label{eq:mix_elbo}
\end{align}
It can be shown that this corresponds to (i) maximizing with respect to $\myvarphi$ a lower bound of the intractable log-marginal likelihood $\ln p_{\myvarphi}(\myx_{1:T})$, and (ii) minimizing with respect to $\myphi'$ the KL divergence between the variational distribution $q_{\myphi'}(\myz_{1:T}|\myx_{1:T})$ and the intractable exact posterior $p_{\mytheta}(\mbf{z}_{1:T} | \mbf{x}_{1:T} )$ \cite{neal1998view}. The proposed VEM algorithm thus consists in iterating between the following variational E and M steps.

\subsubsection{Variational E-step} 

We consider a variational distribution of the same form as the DVAE inference model:
\begin{equation}
    q_{\myphi'}(\myz_{1:T} | \myx_{1:T}) = \prod_{t=1}^T q_{\myphi'}(\myz_{t} | \myz_{1:t-1}, \myx_{1:T}),
    \label{eq:finetune_inference}
\end{equation}
where $ q_{\myphi'}(\myz_{t} | \myz_{1:t-1}, \myx_{1:T})$ is defined as in \eqref{eq:speech_inf}, except that $\mys_t$ is replaced by $\myx_t$. The noisy speech frames can be considered as out-of-sample data for the DVAE model trained on clean speech signals \cite{mattei2018refit}. Therefore, similarly as in \cite{leglaive2020recurrent}, we can fine-tune the pre-trained DVAE inference network on the noisy speech test signal, by maximizing the ELBO in \eqref{eq:mix_elbo} w.r.t.~$\myphi'$. This objective function can be developed by marginalizing and sampling, similarly to what was done in Section~\ref{subsec:b}. This leads to the following expression:
\begin{equation}
    \begin{split}
        &\mcal{L}(\myphi',\myvarphi^{\star};\myx_{1:T}) \overset{c}{=}\, -\sum_{f=1}^{F} \sum_{t=1}^{T} \mbb{E}_{q_{\myphi'}} \left[ \ln v_{\myvarphi^{\star},ft} + \frac{|x_{ft}|^2}{v_{\myvarphi^{\star},ft}} \right] + \\
        &\frac{1}{2} \sum_{l=1}^L \sum_{t=1}^T \Bigg[ \ln v_{\myphi',lt} - \ln v_{\mytheta_{\mbf{z}},lt}  - \frac{v_{\myphi',lt} + ({\mu}_{\myphi',lt} - {\mu}_{\mytheta_{\mbf{z}},lt})^2}{v_{\mytheta_{\mbf{z}},lt}}\Bigg],
        \label{eq:mix_elbo_developped}
    \end{split}
\end{equation}
where $x_{ft}$ denotes the $f$-th entry of $\myx_t$, $\myvarphi^{\star}$ denotes the current estimate of the mixture model parameters, and $v_{\myvarphi^{\star},ft}$ is the $f$-th diagonal entry of $\bs{\Sigma}_{\mytheta_{\mbf{x}},t}$, whose expectation is intractable and is approximated with: 
\begin{equation}
v_{\myvarphi^{\star},ft} = g_t v_{\mytheta_{\mys},ft}(\tilde{\mys}_{1:t-1}, \tilde{\myz}_{1:t}) + (\mbf{W}_b\mbf{H}_b)_{ft}.
\label{var_mix}
\end{equation}
We remind that $\tilde{\mys}_{1:t-1}$ is computed recursively from the output of the decoder network, as explained after equation \eqref{eq:se_method1_approx_s}, and $\tilde{\myz}_{1:t}$ is recursively sampled from $q_{\myphi'}(\myz_{1:t} | \myx_{1:T}) = \prod_{\tau=1}^t q_{\myphi'}(\myz_{\tau} | \myz_{1:\tau-1}, \myx_{1:T})$, as defined in \eqref{eq:finetune_inference}.
During the variational E-step, the parameters $\myphi'$ are updated with a gradient ascent technique, and we denote by $\myphi'^{\star}$ the resulting parameters that will be fixed in the M-step.

We recall that the recursive computation of $\tilde{\mys}_{t}$ is required only for SRNN (actually for the DVAE models with an autoregressive form). DKF and RVAE, as non-autoregressive models, do not require estimating these quantities. They only require the sampling of the latent variables $\tilde{\myz}_{1:t}$.

\subsubsection{M-step} 

The M-step consists in maximizing $\mathcal{L}(\myphi'^{\star}, \myvarphi)$ w.r.t $\myvarphi$ under a non-negativity constraint. Replacing the intractable expectation in \eqref{eq:mix_elbo_developped} with a Monte Carlo estimate (using one single sample), the M-step can be recast as minimizing the following criterion \cite{leglaive2018variance}:
\begin{align}
\mathcal{C}(\myvarphi) &= \sum\limits_{f=1}^{F} \sum\limits_{t=1}^{T} d_{\text{IS}}\left(|x_{ft}|^2, v_{\myvarphi,ft}\right),
\label{eq:mix_Q}
\end{align}
where $v_{\myvarphi,ft}$ is defined in \eqref{var_mix}. This optimization problem can be tackled using a majorize-minimize approach \cite{hunter2004tutorial}, which leads to the multiplicative update rules derived in \cite{leglaive2018variance} using the methodology proposed in \cite{fevotte2011algorithms}:
\begin{equation}
\mathbf{H}_b \leftarrow \mathbf{H}_b \odot \left[ \frac{\mathbf{W}_b^\top \left( | \mathbf{X} |^{\odot 2} \odot \left(\mathbf{V}_\mbf{x} \right)^{\odot -2} \right)}{\mathbf{W}_b^\top  \left(\mathbf{V}_\mbf{x}\right)^{\odot -1} } \right]^{\odot 1/2},
\label{eq:update_H}
\end{equation}
\vspace{.15cm}
\begin{equation}
\mathbf{W}_b \leftarrow \mathbf{W}_b \odot \left[ \frac{\left( | \mathbf{X} |^{\odot  2} \odot \left(\mathbf{V}_\mbf{x}\right)^{\odot -2} \right) \mathbf{H}_b^\top}{ \left(\mathbf{V}_\mbf{x}\right)^{\odot -1} \mathbf{H}_b^\top } \right]^{\odot 1/2},
\label{eq:update_W}
\end{equation}
\vspace{.15cm}
\begin{equation}
\mathbf{g}^\top \leftarrow  \mathbf{g}^\top \odot  \left[ \frac{ \mathbf{1}^\top  \left[| \mathbf{X} |^{\odot 2} \odot \left(\mathbf{V}_\mbf{s} \odot \left(\mathbf{V}_\mbf{x} \right)^{\odot -2}\right)\right]}{\mathbf{1}^\top \left[ \left(\mathbf{V}_\mbf{s} \odot \left(\mathbf{V}_\mbf{x} \right)^{\odot -1}\right)\right]} \right]^{\odot 1/2},
\label{eq:update_g}
\end{equation}
where $\odot$ denotes element-wise multiplication and exponentiation, and matrix division is also element-wise, $\mathbf{V}_\mbf{s}, \mathbf{V}_\mbf{x} \in \mathbb{R}_+^{F \times T}$ are the matrices of entries $v_{\mytheta_{\mbf{s}},ft}$ and $v_{\myvarphi,ft}$ respectively, $\mathbf{X} \in \mathbb{C}^{F \times T}$ is the matrix of entries $x_{ft}$ and $\mathbf{1}$ is an all-ones column vector of dimension $F$. Note that non-negativity is ensured provided that these parameters are initialized with non-negative values.

\subsection{Summary}
In summary, the clean speech signal estimation consists in approximating the posterior $p_{\mytheta} (\mbf{s}_t | \mbf{x}_{1:T})$ and taking the mean of the resulting approximate distribution (i.e., the Wiener filter output). The estimation of the involved parameters is made with the VEM algorithm, which consists in iteratively fine-tuning the inference network of the pre-trained DVAE (E-step) and updating the mixture model parameters $\myvarphi$ (M-step). 
The complete proposed speech enhancement method is summarized in Algorithm \ref{algo:vem-enhancement}. For non-causal DVAEs, we can simply replace $\myz_{1:t}$ with $\myz_{1:T}$ when generating $\mys_{t}$.

\begin{algorithm}[t]
\caption{DVAE-based unsupervised speech enhancement}
\begin{algorithmic}
\Inputs{}\vspace{-4mm}\State{$\triangleright$ Pre-trained DVAE model: $p_{\mytheta} (\mbf{z}_{1:T}, \mbf{s}_{1:T})$ and $q_{\myphi} (\mbf{z}_{1:T} | \mbf{s}_{1:T})$} \State{$\triangleright$ Noisy speech STFT $\mbf{x}_{1:T}$}
\Init{}\vspace{-4mm}\State{$\triangleright$ Initialize NMF noise parameters $\mbf{H}_b$ and $\mbf{W}_b$ with random nonnegative values} 
\State{$\triangleright$ Initialize gain parameters $\mbf{g}= \mbf{1}$}
\State{$\triangleright$ Initialize $q_{\myphi'}(\mbf{z}_{1:t} | \mbf{x}_{1:T} )$ with pre-trained inference network $q_{\bs\phi}(\mbf{z}_{1:t} | \mbf{s}_{1:T} )$}
\While{stopping criterion not reached}
\State{\textbf{E-step:}}
\State{$\triangleright$ Fine-tune $q_{\myphi'}(\mbf{z}_{1:T} | \mbf{x}_{1:T} )$ by maximizing~\eqref{eq:mix_elbo_developped} w.r.t.~$\myphi'$}
\State{$\triangleright$ Sample $\tilde{\mbf{z}}_{1:T}$ from $q_{\myphi'} (\mbf{z}_{1:T} | \mbf{x}_{1:T})$}
\State{$\triangleright$ Compute $\bs{\Sigma}_{\mytheta_{\mys},t}$ for $t=1$ to $T$ using the DVAE decoder}
\State{\textbf{M-step:}}
\State{$\triangleright$ Update $\mbf{H}_b$, $\mbf{W}_b$ and $\mbf{g}$ using (\ref{eq:update_H})-(\ref{eq:update_g})}
\EndWhile
\Outputs{}\vspace{-4mm}
\State{$\triangleright$ Compute the  clean speech signal estimate $\hat{\mbf{s}}_{t}$ for $t=1$ to $T$ using (\ref{eq:se_complete})}
\end{algorithmic}
\label{algo:vem-enhancement}
\end{algorithm}

\section{Experiments}
\label{sec:experiments}

\subsection{Datasets}

We use the WSJ0-QUT dataset and the VoiceBank-DEMAND (VB-DMD) dataset, described below. Each dataset has a ``clean'' version used to pre-train the DVAE models and a ``noisy'' version used to test the proposed speech enhancement algorithm and the reference methods. The clean and noisy versions are actually used together to compute the speech enhancement objective performance measures (see Section~\ref{subsec:metrics}) and for the training of the supervised reference methods. When using only the clean version, we refer to it as WSJ0 or VB.
\subsubsection{WSJ0-QUT} We used the Wall Street Journal dataset (WSJ0) \cite{WSJ0}, which is composed of 16kHz clean speech signals (read Wall Street Journal news). 
WSJ0-QUT is the noisy version
already presented and used in \cite{leglaive2020recurrent}. It was obtained by mixing clean speech signals from WSJ0 with various types of noise signals from the QUT-NOISE dataset \cite{dean2015qut}, with three signal-to-noise ratio (SNR) values: $-5$, $0$, and $5$ dB. The full description of the dataset, including training/test splits and noise types, can be found in \cite{leglaive2020recurrent}. Note that we mixed the speech and noise signals using the ITU-R BS.1770-4 protocol \cite{series2011algorithms}. An SNR computed with this protocol is $2.5$ dB lower (in average) than that computed with sums of squared signal samples.

\subsubsection{VB-DMD} We also used the publicly available VB-DMD dataset \cite{valentini2016investigating}. This dataset contains a training set with $11,572$ utterances performed by $28$ speakers and a test set with $824$ utterances performed by $2$ speakers, different from the training set.
The noisy train set consists of mixture signals mixed at four different SNRs, namely $15$, $10$, $5$, and $0$ dB, whereas the noisy test speech signals are corrupted with $17.5$, $12.5$, $7.5$, and $2.5$ dB SNR and different noise types. The full description of the dataset, including training/test splits and noise types, can be found in \cite{valentini2016investigating}. Following \cite{fu2021metricganU}, we also selected two speakers (p226 and p287) from the clean training set as the validation set for the training of the DVAEs.

\subsubsection{Data preprocessing}
In all our experiments, the STFT was computed with a $64$-ms sine window ($1,024$ samples) and a $75$\%-overlap ($256$ samples hop length), resulting in a sequence of $513$-dimensional discrete Fourier coefficients (for positive frequencies). The DVAEs were trained with STFT power spectrograms of clean speech signals extracted from either WSJ0 or VB, and obtained with the following preprocessing. We first removed the silence at the beginning and ending of the files, using an energy-based voice activity detection threshold of $-30$ dB. The waveform signal was then normalized by its maximum absolute value. We set $T=50$, meaning that speech segments of $0.8$s were used to train the DVAE models. In summary, each training data sequence is a $513 \times 50$ STFT power spectrogram. For WSJ0, this data preprocessing resulted in a set of $N_{\rm{tr}} = 93,393$ training sequences (representing about $20.8$ hours of speech signal) and $N_{\rm{val}} = 7,775$ validation sequences (about $1.7$ hours). For VB, we obtained $N_{\rm{tr}} = 29,389$ training sequences ($6.5$ hours) and $N_{\rm{val}} = 2,152$ validation sequences ($0.5$ hour). For the evaluation of the speech enhancement methods, we used the STFT spectrogram of each complete noisy test sequence (with normalization), which can be of variable length, most often larger than $2$s. For WSJ0-QUT, the total duration of the test dataset is $1.5$ hours, and for VB-DMD it is $0.6$ hour.
 
\begin{figure*}[htbp]
    \centering
    \begin{tabular}{ccc}
    \includegraphics[width=0.29\linewidth]{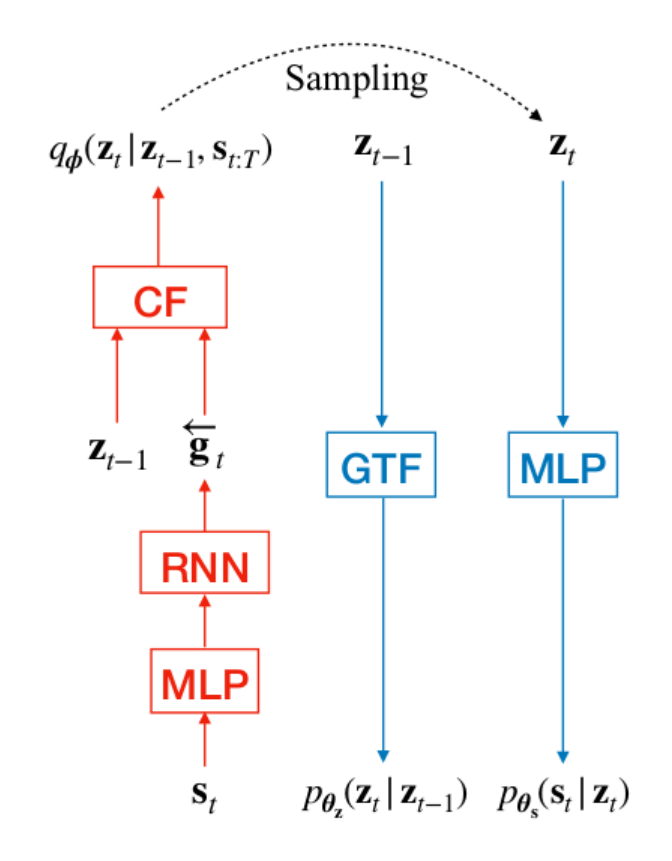} &
    \includegraphics[width=0.28\linewidth]{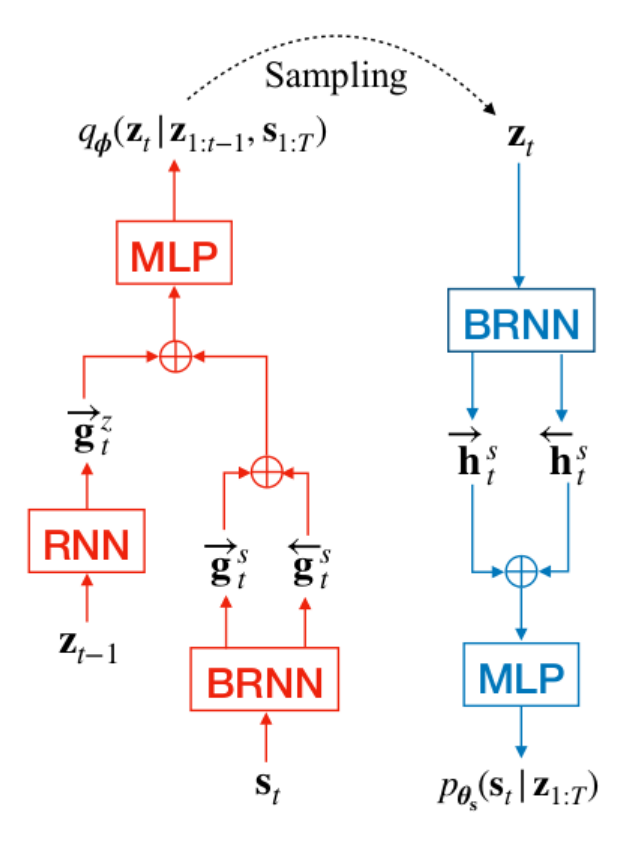} &
    \includegraphics[width=0.41\linewidth]{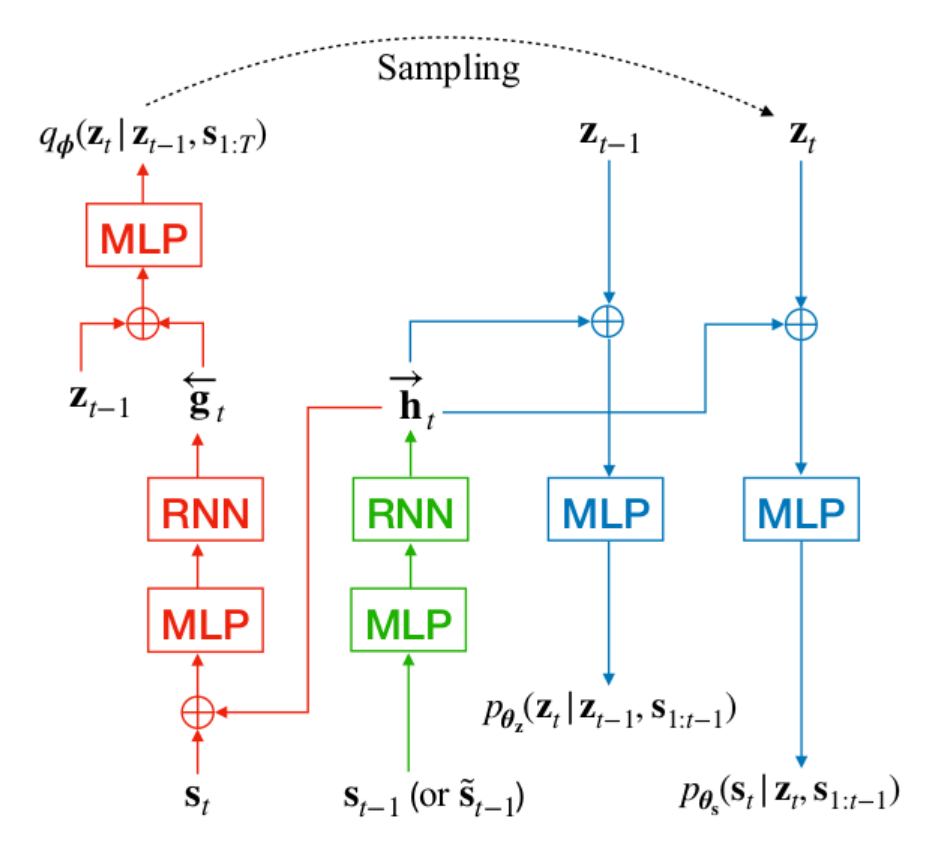} \\
    {(a) DKF} & {(b) RVAE} & {(c) SRNN}
    \end{tabular}
    \caption{Model implementation for the (a) deep Kalman filter (DKF)\cite{krishnan2015deep,krishnan2017structured}, (b) recurrent variational autoencoder (RVAE) \cite{leglaive2020recurrent}, and (c) stochastic recurrent neural network (SRNN) \cite{fraccaro2016sequential}. Each model consists of an inference (encoder) network (in red) and a generation (decoder) network (in blue). SRNN has a shared module between encoder and decoder (in green). CF: combiner function, GTF: gated transition function (see \cite{krishnan2015deep} for details), RNN: recurrent neural network, BRNN: bidirectional RNN, MLP: multi-layer perceptron, {\scriptsize{$\bigoplus$}}: vector concatenation. All RNNs are implemented with LSTM networks.}
    \label{fig:arc}
\end{figure*}

\subsection{Evaluation metrics} 
\label{subsec:metrics}

We used three metrics to evaluate the quality of the estimated speech signals: The scale-invariant signal-to-distortion ratio (SI-SDR) in dB \cite{le2019sdr}, the perceptual evaluation of speech quality (PESQ) score~\cite{rix2001perceptual}, and the extended short-time objective intelligibility (ESTOI) score (in $[0, 1]$) \cite{taal2011algorithm}. The PESQ measure is declined in three different variants, depending on different protocols:\footnote{See the explanation for `pesq' at \url{https://github.com/ludlows/python-pesq} and for `pypesq' and \url{https://github.com/vBaiCai/python-pesq}. } the narrow-band PESQ MOS value (PESQ MOS, in $[-0.5, 4.5]$), the narrow-band PESQ LQ0 value (PESQ NB, in $[1, 5]$), and the wide-band PESQ LQ0 value (PESQ WB, in $[1, 5]$). We report all of them in the following experiments. For all measures, a higher value indicates a better result.

\subsection{Models implementation}

Here we present the implementation of the three example DVAEs that we used in practice in the proposed DVAE-based speech enhancement algorithm, namely DKF, RVAE, and SRNN. As indicated in \cite{girin2020dynamical}, we can have various implementations for each DVAE model, thus we only present the model configurations that showed the best performance in our experiments (for the latent space dimension selected below).

\subsubsection{Dimension of the latent space}
In the present experiments, we set $L=16$. We recall that the data dimension is $F=513$. We also recall that $\myz_t$ is a real-valued vector that is modeled by a Gaussian distribution, so the DNNs modeling $\myz_t$ has to output two $L$-dimensional vectors, the mean and variance vectors, for both inference and generation (except for RVAE since $\myz_t$ is assumed i.i.d. with a standard Gaussian distribution and no DNN is used for its generation). In contrast, $\mys_t$ is a complex-valued vector modeled by a circular complex Gaussian distribution, which only leaves one $F$-dimensional variance vector to be provided by the decoder DNN. To guarantee the positivity of the entries of this output variance vector, we used log-parameterization (the output is the log-variance in $\mathbb{R}$, which is then converted to variance by taking the exponential). The last layer predicting the mean and log-variance parameters is always a linear layer, with a dimension corresponding to that of $\myz_t$ (16) or $\mys_t$ (513). We omit this in the following description for simplicity.

\subsubsection{DKF}
Fig.~\ref{fig:arc}(a) summarizes our implementation of DKF. The layers providing the parameters of the inference and generative models of $\myz_t$ are respectively implemented with the specific \textit{combiner function} and \textit{gated transition function} described in \cite{krishnan2017structured}.
For the inference model, we used a backward long short-term memory (LSTM) \cite{hochreiter1997long} layer with an internal state of dimension $128$ to accumulate the information from $\mys_{t:T}$ in (\ref{eq:dkf_inf}). Before being fed into the recurrent layer, each vector $\mys_t$ passes through a multi-layer perceptron (MLP) with one hidden layer of dimension $256$ and a $\tanh$ activation. The variance parameters of the generative model of $\mys_t$ are provided by an MLP with 4 hidden layers of dimension $32$, $64$, $128$ and $256$, with a $\tanh$ activation function. 

\subsubsection{RVAE} We implemented the non-causal version of RVAE as schematized in
Fig.~\ref{fig:arc}(b). The inference model includes a bidirectional LSTM (BLSTM) layer with an internal state of dimension $128$ to process the complete sequence $\mathbf{s}_{1:T}$ and an LSTM layer to process the sampled past latent vectors sequence $\myz_{1:t-1}$. The output of these two layers are then concatenated and mapped into the parameters of the inference model over $\myz_t$ by an MLP.
The generative part of the model includes a BLSTM layer with an internal state of dimension 128, which takes the sampled $\myz_{1:T}$ as input. The output of this BLSTM layer is finally mapped to the parameters of the generative model over $\mys_t$ by a single linear layer.

\subsubsection{SRNN}
SRNN is quite different from the two previous models. As shown in Fig.~\ref{fig:arc}(c), the inference and generative models share a recurrent internal state vector $\overrightarrow{\bs{h}}_t$ (module in green) that is encoding the information from the past observed vectors $\mys_{1:t-1}$.
This shared module is composed of an MLP with one layer of dimension $256$ followed by a forward LSTM. The dimension of $\overrightarrow{\bs{h}}_t$ is $128$. For inference, the concatenation of $\overrightarrow{\bs{h}}_t$ and $\mys_t$ is fed into a one-layer MLP of dimension $256$ followed by a backward LSTM that provides the vector $\overleftarrow{\mbf{g}}_t$ (of dimension $128$). This vector is then concatenated with the sample of $\myz_{t-1}$ and fed into an MLP with two hidden layers of dimension $64$ and $32$. For the generative part, we concatenate the shared state $\overrightarrow{\bs{h}}_t$ with the sampled latent vector at the previous or current time frame. An MLP with two hidden layers of dimension $64$ and $32$ is used for the generation of $\myz_t$, and an MLP with one hidden layer of dimension 128 is used for the generation of $\mys_t$. All MLP hidden layers use the $\tanh$ activation function.

\subsection{DVAEs pre-training}

For the pre-training of the three DVAE models on clean speech signals, we used the Adam optimizer \cite{kingma2014adam} with a learning rate of $1e\!-\!3$ and $\beta_1=0.9$, $\beta_2=0.99$. On both datasets, we trained each model with a batch size of $128$ during $300$ epochs and kept the model snapshot with lowest validation loss. We applied a linear KL annealing for the first $20$ epochs to warm-up the latent space \cite{sonderby2016ladder}.

As an autoregressive model, SRNN deserves a particular treatment during pre-training. Indeed, in the conventional training of autoregressive models, the ground-truth past clean speech vectors $\mys_{1:t-1}$ are used to generate the current one $\mys_{t}$, a strategy sometimes referred to as ``teacher forcing'' in the literature \cite{williams1989learning}. We have seen in Section~\ref{subsubsec:pastSpeechVectMarg} that it is not possible to do that in the proposed speech enhancement algorithm, where $\mys_{1:t-1}$ is replaced by its proxy $\tilde{\mbf{s}}_{1:t-1}$ (recursively computed from the decoder output). It is shown in \cite{girin2020dynamical} that directly using $\tilde{\mbf{s}}_{1:t-1}$ in the SRNN model trained with $\mys_{1:t-1}$ significantly decreases the quality of the reconstructed speech spectrogram, due to the mismatch between train and test conditions. To avoid such a mismatch (here between DVAE training and speech enhancement conditions), we trained SRNN using $\tilde{\mbf{s}}_{1:t-1}$ (instead of $\mys_{1:t-1}$) to generate $\mys_{t}$. Such a training is difficult in practice and to make it efficient, we adopted a ``scheduled sampling'' approach \cite{bengio2015scheduled}, i.e., we progressively replace $\mys_{1:t-1}$ with $\tilde{{\mys}}_{1:t-1}$, with a proportion going from $0\%$ to $100\%$ along the training iterations. That is, at the beginning of the training, $\mys_t$ is generated completely from $\mys_{1:t-1}$ and $\myz_{t}$. Then the probability to use $\tilde{{\mys}}_{1:t-1}$ increases during the training procedure. Finally, $\mys_t$ is generated completely from $\tilde{\mys}_{1:t-1}$ and $\myz_{t}$. This takes 80 epochs after the KL annealing step.


\begin{table}[t]
\caption{Results of the speech analysis-resynthesis experiment, averaged over the test subset of WSJ0  and VoiceBank.}\vspace{-2mm}
\label{tab:dvae_resyn}
\centering
\resizebox{\columnwidth}{!}{\setlength{\tabcolsep}{1.2mm}{
\begin{tabular}{l c c c c c c}
\toprule
Models & Dataset & SI-SDR (dB) & PESQ MOS & PESQ WB & PESQ NB & ESTOI \\
\midrule
VAE & WSJ0 & 8.0 & 3.33 & 2.95 & 3.31 & 0.89 \\
DKF & WSJ0 & 9.0 & 3.55 & 3.39 & 3.61 & 0.91  \\
RVAE & WSJ0 & 9.8 & 3.65 & 3.57 & 3.75 & 0.92 \\
SRNN & WSJ0 & 8.2 & 3.48 & 3.24 & 3.52 & 0.90\\
\midrule
VAE & VB & 8.6 & 3.22 & 2.79 & 3.15 & 0.88 \\
DKF & VB & 9.4 & 3.35 & 2.96 & 3.34 & 0.90 \\
RVAE & VB & 9.6 & 3.41 & 3.00 & 3.42 & 0.90 \\
SRNN & VB & 9.1 & 3.39 & 2.99 & 3.39 & 0.89\\
\bottomrule
\end{tabular}}}
\end{table}

Before we examine the speech enhancement performance, we can rapidly compare the speech modeling capacities of the three selected DVAE models (and the vanilla VAE) after their pre-training, by conducting a speech analysis-resynthesis experiment (i.e., chaining of the encoder and decoder) similar to~\cite{bie21_interspeech}. The overall pipeline is shown in Fig.~\ref{fig:resynthesis}. For the VAE model, we used the baseline architecture already used in \cite{leglaive2020recurrent}. The results presented in Table~\ref{tab:dvae_resyn} were obtained with the models being trained on the WSJ0 or VB train subsets and averaged over the corresponding test subsets. For SI-SDR scores, the noise is the modeling noise, i.e., the difference between original and reconstructed signal. We can see from Table~\ref{tab:dvae_resyn} that all DVAE models outperform the VAE for all metrics, both on WSJ0 and VB, showing the benefits of introducing dynamics into VAE-based speech modeling.

\begin{table*}[t!]
\caption{Speech enhancement results obtained with models trained and tested on corresponding datasets. * indicates the implementation provided by the authors and re-trained. The ``Supervision'' column indicates whether the training is supervised (S), unsupervised noise-dependent (UD) or noise-agnostic (UA).}\vspace{-2mm}
\label{tab:dvae_se_uni_dataset}
\centering
\setlength{\tabcolsep}{1.2mm}{
\begin{tabular}{l | cccc | ccccc}
\toprule
Method & Supervision & Parameters & Train subset & Test subset & SI-SDR (dB) & PESQ MOS & PESQ WB & PESQ NB & ESTOI \\
\midrule
Noisy mixture & - & - & - & WSJ0-QUT 
& $-$2.6 & 1.83 & 1.14 & 1.57 & 0.50 \\
\midrule
VAE-VEM~\cite{leglaive2020recurrent} & UA & 0.14M & WSJ0 & WSJ0-QUT &
5.0 & 2.13  & 1.45  &  1.86 & 0.58\\
Proposed DKF-VEM & UA & 0.52M & WSJ0 & WSJ0-QUT &
5.1	& 2.23 & 1.46 & 1.95 & 0.62 \\
Proposed RVAE-VEM & UA & 1.06M & WSJ0 & WSJ0-QUT & 
\textbf{5.8} & \textbf{2.27} & \textbf{1.54} &\textbf{ 1.98} & 0.62 \\
Proposed SRNN-VEM & UA & 0.88M & WSJ0 & WSJ0-QUT & 
5.2 & 2.23 & 1.48 & 1.95 & \textbf{0.63} \\

\midrule
UMX*~\cite{uhlich2020open} & S & 1.55M & WSJ0-QUT & WSJ0-QUT & 
\textbf{5.7} & 2.16 & 1.38 & 1.83 & \textbf{0.63}\\
MetricGAN+*~\cite{fu2021metricgan+} & S & 1.90M & WSJ0-QUT  & WSJ0-QUT & 
3.6 & \textbf{2.83} & \textbf{2.18} & \textbf{2.61} & 0.60 \\

\midrule
\midrule
Noisy mixture & - & - & - & VB-DMD
& 8.4 & 3.02 & 1.97 & 2.88 & 0.79 \\
\midrule
NyTT~\cite{fujimura2021noisy} & UD & - & VB-DMD + Extra noise &  VB-DMD &
\textbf{17.7} & - & 2.30 & - & - \\
NyTT~\cite{fujimura2021noisy} & UD & - & VB-DMD & VB-DMD & 
12.1 & - & 1.74 & - & - \\
MetricGAN-U (full)~\cite{fu2021metricganU} & UD & 1.90M & VB-DMD  & VB-DMD & 
6.5  & 3.13  & 2.13  & 3.03  & 0.74 \\
MetricGAN-U (half)~\cite{fu2021metricganU} & UD & 1.90M & VB-DMD  & VB-DMD & 
8.2  & 3.20  & 2.45  & 3.11  & 0.77 \\
VAE-VEM~\cite{leglaive2020recurrent} & UA & 0.14M & VB & VB-DMD & 
16.4 & 3.18 & 2.37 & 3.10 & 0.80  \\
Proposed DKF-VEM & UA & 0.52M & VB & VB-DMD & 
16.9 & 3.22 & 2.42 & 3.14 & \textbf{0.81}  \\
Proposed RVAE-VEM & UA & 1.06M & VB & VB-DMD & 
17.1 & \textbf{3.23} & \textbf{2.48} & \textbf{3.15} & \textbf{0.81}  \\
Proposed SRNN-VEM & UA & 0.88M & VB & VB-DMD & 
14.2 & 3.20 & 2.32 & 3.12 & 0.80\\
\midrule
UMX~\cite{uhlich2020open}  & S & 1.55M & VB-DMD & VB-DMD & 
\textbf{14.0} & 3.18 & 2.35 & 3.08 & \textbf{0.83}  \\
MetricGAN+~\cite{fu2021metricgan+} & S & 1.90M  & VB-DMD  & VB-DMD & 
8.5 & \textbf{3.59} & \textbf{3.13} & \textbf{3.63} & \textbf{0.83}  \\

\bottomrule
\end{tabular}
}
\end{table*}

\subsection{DVAE-VEM algorithm settings}
\begin{figure}[t]
    \centering
    \begin{tabular}{r}
    \includegraphics[width = 0.97\linewidth]{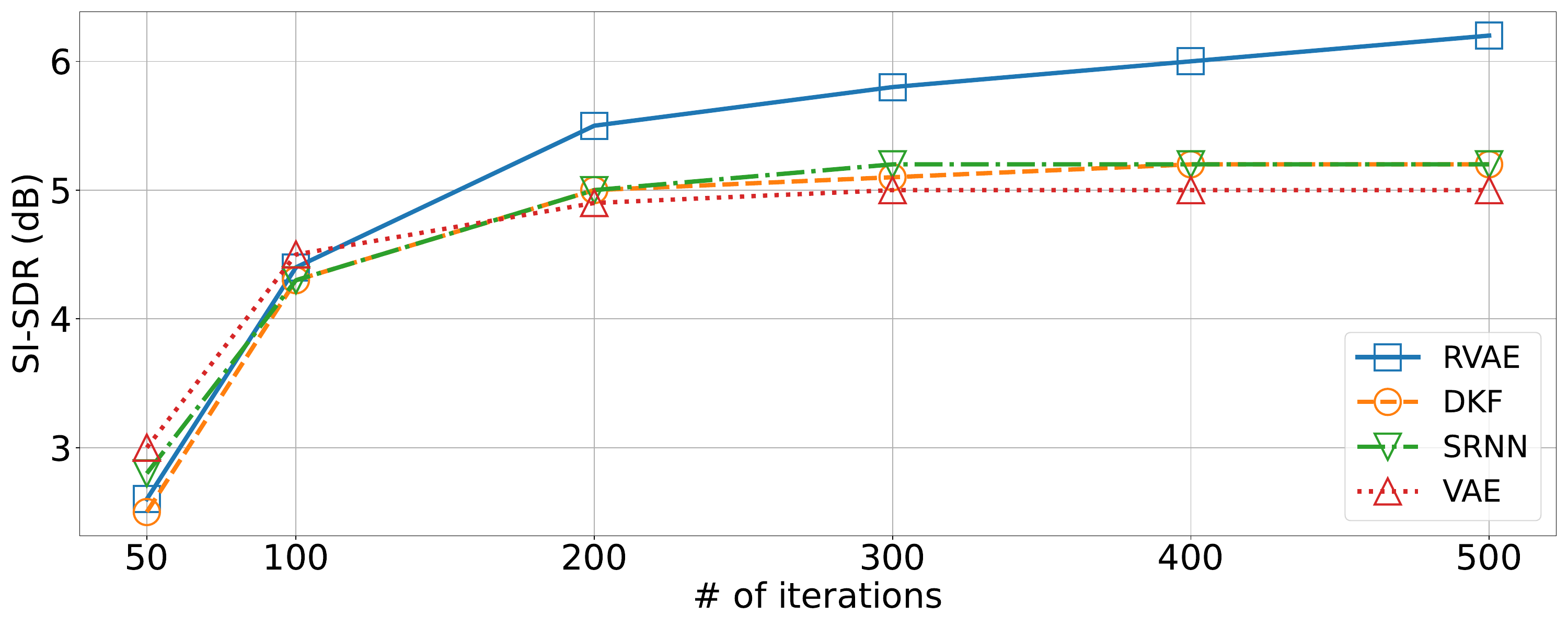}\\
    \includegraphics[width = 0.985\linewidth]{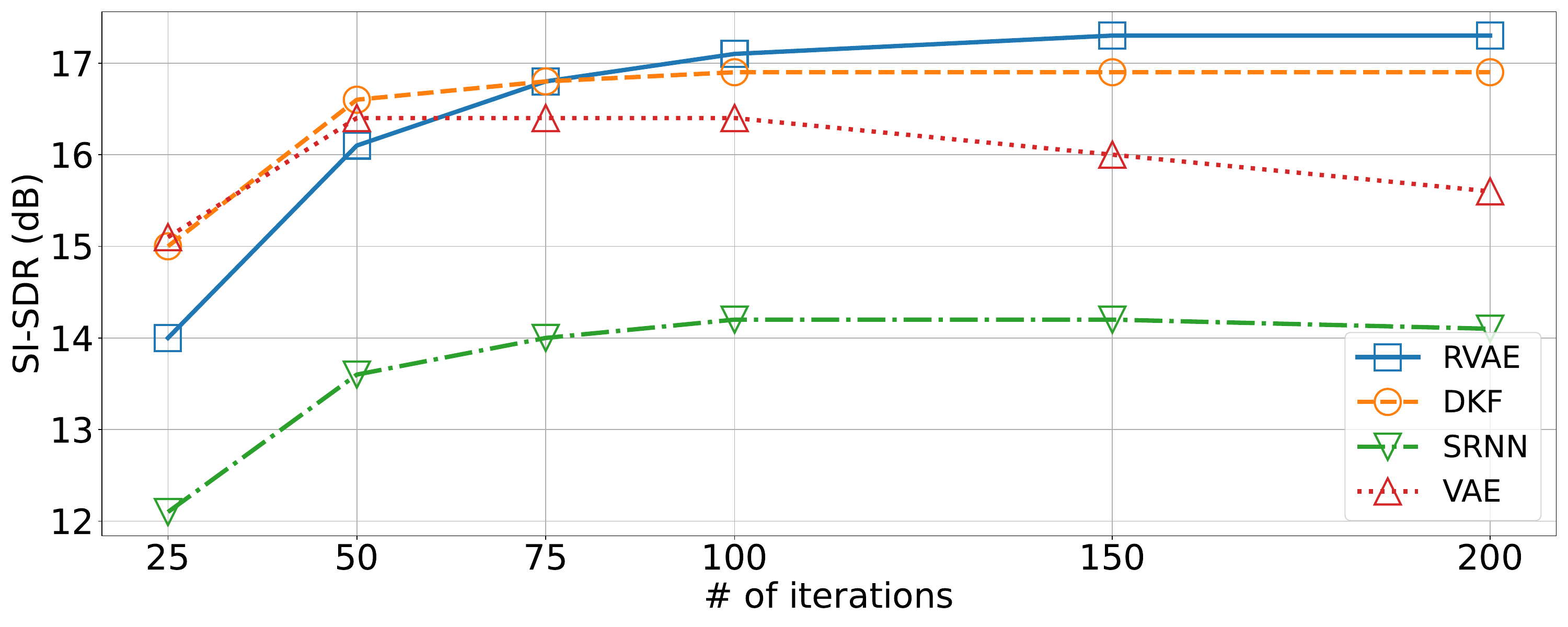}\\
    \end{tabular}
    \caption{Performance of the proposed DVAE-based speech enhancement method (and VAE baseline) as a function of the number of VEM iterations, for the WSJ0-QUT (top) and VB-DMD (bottom) test datasets. All DVAE models are pre-trained on the corresponding dataset of clean speech signals (WSJ0 and VB, respectively).}
    \label{fig:iter}
\end{figure}

The rank of the NMF in the noise model \eqref{eq:nmf_noise} is set to $K=8$. $\mathbf{W}_b$ and $\mathbf{H}_b$ were randomly initialized from a uniform distribution in $[0,1]$ and $\mathbf{g}$ was initialized with an all-ones vector. In the E-step of the VEM algorithm, the encoder of the DVAE models is fine-tuned using the Adam optimizer~\cite{kingma2014adam} with a learning rate of $1e\!-\!3$.  Fig.~\ref{fig:iter} illustrates the influence of the number of VEM iterations on the performance of the different DVAE models on the two test datasets. We observe that the performance of most models plateaus from 300 and 100 iterations for WSJ0-QUT and VB-DMD respectively. We fixed the number of iterations to these values as they represent a global optimal trade-off between performance and complexity and are neither beneficial nor disadvantageous for a particular DVAE model.

\subsection{Baselines}
\label{subsec:baselines}

Regarding baseline supervised methods, we used Open-Unmix (UMX) and MetricGAN+. UMX is an open-source method based on a BLSTM network. It was originally proposed for music source separation \cite{stoter2019open} and was later adapted for speech enhancement \cite{uhlich2020open}. MetricGAN+ \cite{fu2021metricgan+} also adopts BLSTMs for mask-based prediction of the clean speech. In addition, it introduces a metric network that is trained to approximate the PESQ evaluation score, and then MetricGAN+ is trained to maximise this proxy of the PESQ score.

Regarding baseline unsupervised methods, we choose VAE-VEM \cite{leglaive2020recurrent}, MetricGAN-U \cite{fu2021metricganU} and NyTT \cite{fujimura2021noisy}. VAE-VEM, referred to as VAE-FFNN in ~\cite{leglaive2020recurrent}, uses the same optimization methodology as our approach, except there is no temporal model. In practice, we tried with different model complexities for the VAE, and we report the one exhibiting the best performance. MetricGAN-U is the unsupervised version of MetricGAN+. Since the supervision in MetricGAN+ comes from the PESQ score computation using the paired enhanced and clean speech signal, MetricGAN-U adopts a non-intrusive speech quality metric instead, namely the DNSMOS measure \cite{reddy2021dnsmos}, to bypass the paired supervision. Two versions of MetricGAN-U were proposed by the authors: the ``full'' one is trained entirely without supervision, while the ``half'' version monitors the PESQ measure to perform early-stopping. Since PESQ is an intrusive measure, this version of MetricGAN-U can be seen as weakly supervised (supervision is used only for validation and not for training \textit{per se}) \cite{fu2021metricganU}. NyTT is based on a noisy speech target training strategy, where the network is trained to remove an additional noise added to the noisy speech. Since there is no need for noisy/clean speech pairs, NyTT can also be considered as an unsupervised speech enhancement method. It should be noted that both MetricGAN-U and NyTT are trained using noisy speech as input, so the resulting model is noise-dependent. This contrasts with the proposed method where only clean speech is used for the DVAE pre-training, resulting in a noise-agnostic speech enhancement method. Finally, we could not re-train MetricGAN-U or NyTT. Indeed, since MetricGAN-U uses the DNSMOS service, each training epoch needs 2 days for evaluation, which is impractical. Regarding NyTT, we cannot retrain it since there is no public release of the code.

\subsection{Speech enhancement results}
\label{subsec:results-SE}

The speech enhancement scores obtained with the proposed method (for the three DVAE models) and with the baseline methods are reported in Tables~\ref{tab:dvae_se_uni_dataset} and \ref{tab:dvae_se_cross_dataset}, along with the capacity of the models, reported in terms of number of parameters.
Table~\ref{tab:dvae_se_uni_dataset}
shows the results obtained when the test subset ``corresponds'' to the train subset, i.e., it originates from the same dataset (WSJ0 or WSJ0-QUT in the upper half of the table, and VB or VB-DMD in the lower half). Table~\ref{tab:dvae_se_cross_dataset} shows the results obtained with cross-dataset experiments conducted to evaluate the generalization capability of the different models. This means that WSJ0 or WSJ0-QUT is used for training and VB-DMD is used for testing, or alternatively, VB or VB-DMD is used for training and WSJ0-QUT is used for testing. Baseline models marked with * in Table~\ref{tab:dvae_se_uni_dataset} were retrained using the implementation provided by the authors. Other baseline results are obtained from the corresponding papers or from the pre-trained models if available. It can be seen that the number of parameters for the three DVAE models is lower than that of all the baselines but the VAE-VEM method. Among the DVAE models, RVAE is the one with the highest number of parameters (1.06M for RVAE, 0.88M for SRNN and 0.52M for DKF).

From the results in Table~\ref{tab:dvae_se_uni_dataset}, we first observe that the proposed DVAE-VEM algorithm outperforms the VAE-based counterpart for all three DVAE models, except SRNN-VEM on the VB dataset. This is consistent with the results of the analysis-resynthesis experiment and this shows the interest of modeling the speech signal dynamics within the proposed speech enhancement method. Among the three tested DVAE models, RVAE performs the best for all evaluation metrics, except in terms of ESTOI on the WSJ0-QUT dataset, where SRNN obtains a slightly better score. 

\begin{table*}[t!]
\caption{Speech enhancement results with models trained and tested on different datasets. The ``Supervision'' column indicates whether the training is supervised (S), unsupervised noise-dependent (UD) or noise-agnostic (UA).}\vspace{-2mm}
\label{tab:dvae_se_cross_dataset}
\centering
\resizebox{0.9\textwidth}{!}{\setlength{\tabcolsep}{1.2mm}{
\begin{tabular}{l | cccc | ccccc}
\toprule
Method & Supervision & Parameters & Train subset & Test subset & SI-SDR (dB) & PESQ MOS & PESQ WB & PESQ NB & ESTOI \\
\midrule
Noisy mixture & - & - & - & WSJ0-QUT 
& $-$2.6 & 1.83 & 1.14 & 1.57 & 0.50 \\
\midrule
MetricGAN-U (full)~\cite{fu2021metricganU} & UD & 1.90M   & VB-DMD & WSJ0-QUT & 
$-$2.3 & 1.91 & 1.18 & 1.63 & 0.50 \\
MetricGAN-U (half)~\cite{fu2021metricganU} & UD & 1.90M  & VB-DMD & WSJ0-QUT & 
$-$1.6 & 2.01 & 1.25 & 1.71 & 0.49\\
VAE-VEM~\cite{leglaive2020recurrent} & UA & 0.14M & VB & WSJ0-QUT & 
3.8 & 1.89 & 1.31 & 1.68 & 0.54  \\
Proposed DKF-VEM & UA & 0.52M  & VB & WSJ0-QUT & 
3.5 & 2.08 & 1.32 & 1.80 & 0.57 \\
Proposed RVAE-VEM & UA & 1.06M  & VB & WSJ0-QUT & 
4.3 & 2.12 & 1.37 & 1.84 & 0.57  \\
Proposed SRNN-VEM & UA & 0.88M & VB & WSJ0-QUT & 
\textbf{4.6} & \textbf{2.21} &\textbf{ 1.42} & \textbf{1.91} & \textbf{0.61} \\

\midrule
UMX~\cite{uhlich2020open} & S & 1.55M & VB-DMD & WSJ0-QUT & 
\textbf{4.1} & 2.06 & 1.34 & 1.76 & \textbf{0.61} \\
MetricGAN+~\cite{fu2021metricgan+} & S & 1.90M & VB-DMD & WSJ0-QUT & 
1.8 & \textbf{2.31} & \textbf{1.61} & \textbf{2.02} & 0.56  \\

\midrule
\midrule

Noisy mixture & - & - & - & VB-DMD
& 8.4 & 3.02 & 1.97 & 2.88 & 0.79 \\
\midrule

VAE-VEM~\cite{leglaive2020recurrent} & UA & 0.14M & WSJ0 & VB-DMD & 
15.0 & 3.16 & 2.27 & 3.06 & 0.79 \\
Proposed DKF-VEM & UA & 0.52M & WSJ0 & VB-DMD & 
16.8 & 3.17 & 2.34 & 3.08 & \textbf{0.81} \\
Proposed RVAE-VEM & UA & 1.06M & WSJ0 & VB-DMD & 
\textbf{17.3} & \textbf{3.21} & \textbf{2.41} & \textbf{3.13} & \textbf{0.81} \\
Proposed SRNN-VEM & UA & 0.88M  & WSJ0 & VB-DMD & 
16.8 & 3.17 & 2.34 & 3.08 & \textbf{0.81} \\

\midrule
UMX*~\cite{uhlich2020open} & S & 1.55M  & WSJ0-QUT & VB-DMD & 
\textbf{10.4} & 3.10 & 2.21 & 2.98 & \textbf{0.78}  \\
MetricGAN+*~\cite{fu2021metricgan+} & S & 1.90M & WSJ0-QUT & VB-DMD & 
3.9 & \textbf{3.41} & \textbf{2.51} & \textbf{3.39} & 0.73 \\
\bottomrule
\end{tabular}
}}
\end{table*}

When comparing with the baseline unsupervised methods, hence only on the VB-DMD dataset, the proposed method achieves competitive results. Even if the highest SI-SDR performance is achieved by NyTT ($17.7$~dB), this is possibly due to the use of large amounts of noise, given that its performance drops significantly ($12.1$~dB) when only the noise from the DMD dataset is used. The proposed RVAE-VEM algorithm reaches very similar performance ($17.1$~dB) without training on any kind of noise. In terms of PESQ WB, the proposed approaches outperform NyTT independently of the amount of noise used at training time. Similar conclusions are drawn when comparing to MetricGAN-U, in which case the performance difference in terms of SI-SDR is very large ($+8$ to $+10$~dB). Remarkably, the proposed method achieves competitive performance with, and sometimes outperforms, MetricGAN-U (half) in terms of PESQ WB score, even though MetricGAN-U (half) uses the PESQ WB score on the validation set as a training stop criterion. In this regard, MetricGAN-U (half) is expected to exhibit higher PESQ WB values than MetricGAN-U (full).

When comparing with the supervised methods, we see that MetricGAN+ obtains PESQ scores that are significantly higher (e.g., $3.59$ and $2.83$ PESQ NB on VB-DMD and WSJ0-QUT respectively) than those of all the other methods (RVAE-VEM reaches a maximum of $3.23$ and $2.27$ on VB-DMD and WSJ0-QUT). This can be explained by the fact that PESQ is the criterion optimized during the MetricGAN+ model training. In contrast, when it is evaluated with other metrics, the results obtained by this method are considerably worse, especially in terms SI-SDR for which MetricGAN+ performs the worst. We found that this may be due to the fact that the energy in the speech signal estimated by MetricGAN+ is mostly concentrated in the low-frequency part of the spectrum, whereas the mid- and high-frequency parts are poorly recovered. Regarding UMX, we observe that its performance is systematically under the one of the proposed method, in terms of SI-SDR or any of the PESQ measures for both datasets.

To evaluate the generalization capability of the different models, we report in Table~\ref{tab:dvae_se_cross_dataset} their performance obtained when the training set and test set originate from different corpora. It is unsurprising that the performance of supervised methods significantly decreases in this setting. For example, MetricGAN+ obtains a PESQ WB value of $3.13$ on the VB-DMD test dataset when trained on the VB-DMD train dataset, whereas it drops to $2.51$ when trained on the WSJ0-QUT. In the same vein, UMX goes from $14.0$ dB SI-SDR to $10.4$ dB SI-SDR. Similary behavior is found for the other metrics and datasets. As for the unsupervised noise-dependent methods, we can also see a significant decrease of performance. For example, Metric-GAN-U (trained on VB-DMD and provided by the authors) obtains a negative SI-SDR when tested on WSJ0-QUT. In contrast, the proposed DVAE-VEM algorithm is much less affected by the dataset mismatch (for all DKF, RVAE, and SRNN models). Even if, generally speaking, we observe a mild decrease in performance when the training and test sets do not match, the proposed method occasionally exhibits better performance when tested in a different dataset. For example, the RVAE-VEM algorithm tested on VB-DMD with RVAE trained on WSJ0 obtains $17.3$ dB of SI-SDR, which is even better than when RVAE is trained on VB ($17.1$ dB). This is probably because the WSJ0 dataset is larger than the VB dataset. RVAE-VEM performs the best when trained on WSJ0 and tested on VB-DMD, while SRNN-VEM performs the best when trained on VB and tested on WSJ0-QUT. Overall, we can conclude that the proposed DVAE-based speech enhancement method exhibits competitive performance when compared to other supervised and unsupervised methods in the ``corresponding dataset'' setting, and superior performance in the cross-dataset setting.

\begin{figure}[t]
    \centering
    \includegraphics[width = 1.03\linewidth]{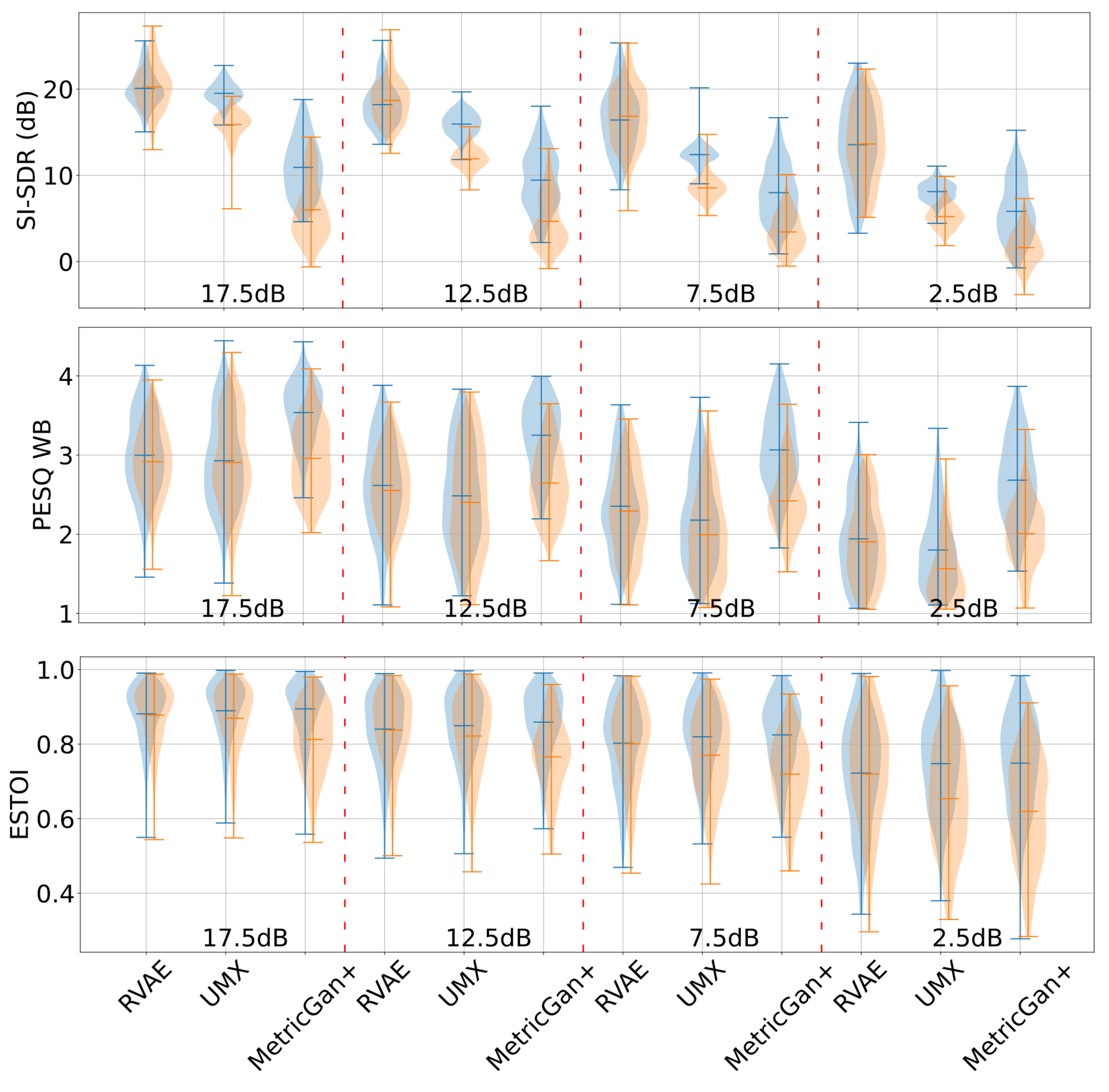}
    \caption{Detailed performance of RVAE-VEM, UMX and MetricGAN+ on the VB-DMD dataset for different input SNRs. The blue and orange violin plots correspond to a training of the models on the VB-DMD and the WSJ0-QUT datasets, respectively.}
    \label{fig:compare}
\end{figure}

Complementary to the above experimental analysis, we present in Fig. \ref{fig:compare} violin plots showing the full distribution of the results obtained with RVAE-VEM, UMX and MetricGAN+, when evaluated on the VB-DMD dataset and trained either on the corresponding training set or on the WSJ0-QUT dataset. The results are presented separately for different test SNRs. As expected, we observe that the performance of all methods degrades as the SNR decreases.
We can also see that the proposed method (RVAE) is much less affected by the mismatch between training and test sets, compared with supervised methods (UMX and MetricGAN+) for which the distributions of the results clearly shift down.

At the light of the presented results, we extract the following concluding remarks. When there is no mismatch between the training and test datasets, the proposed method achieves state-of-the-art performance in unsupervised speech enhancement, and competitive results when compared with supervised speech enhancement methods (often outperforming them depending on the metric). As for cross-dataset experiments, where the training and test sets are coming from different datasets, we observe that the performance of the supervised methods is severely affected by the dataset mismatch, whereas the performance of the proposed unsupervised method is very robust to it. Overall, the results obtained in the various settings demonstrate the interest of the proposed DVAE-VEM methodology for speech enhancement. Audio examples and code are available at \url{https://team.inria.fr/robotlearn/unsupervised-speech-enhancement-using-dynamical-variational-auto-encoders}.

\section{Conclusion}
\label{sec:conclusion}

In this paper, we have proposed a general framework for unsupervised speech enhancement based on DVAEs. In our framework, the DVAEs are used to model the clean speech signal, while the noise is modeled via NMF. While DVAEs are pre-trained with a clean speech dataset, the noise parameters are estimated at test time, together with the clean speech, from the noisy speech sequence to process. To achieve that, we have derived a VEM algorithm for the most general formulation of a DVAE model, which can then be easily adapted to particular instances of DVAEs. We have illustrated this principle with DKF, RVAE and SRNN, and this can be extended to other DVAE models, e.g., STORN \cite{bayer2014learning} or VRNN \cite{chung2015recurrent}.

We have evaluated the speech enhancement performance obtained with those three example DVAEs. The proposed approach exhibits superior or competitive performance compared to supervised and unsupervised state-of-the-art methods when the training and test datasets are from the same corpora, and outperforms them on cross-dataset settings, i.e., when the training and test datasets are from different corpora. The RVAE model provided the best performance among the tested DVAEs. SRNN shows a great potential, provided that it is trained with scheduled sampling in order to reduce the gap between the training and speech enhancement conditions. If this gap could be further decreased, we believe that it could have even better performance than RVAE. This aspect should be further investigated, possibly including other autoregressive models in the DVAE family (e.g., VRNN \cite{chung2015recurrent}). 

So far, the good performance of the proposed iterative DVAE-VEM algorithm comes at the cost of a high computational time. Indeed, processing one second of audio with 100 iterations of the algorithm takes approximately 14, 25 and 21 seconds for DKF, RVAE and SRNN, respectively, using a single core of an Intel Xeon Gold 6230 at 2.1GHz. Future work will include developing fast DVAE-based speech enhancement algorithms, for instance inspiring from \cite{pariente2019statistically}. Also, so far, the inference with DVAEs is non-causal, meaning that past, present, and future noisy speech observations are required to enhance a given speech frame. Causal DVAE-based speech enhancement can also be investigated, but is out of the scope of this paper. Future work also includes using other powerful encoder-decoder networks, e.g., TCNs \cite{lea2016temporal} and the Transformer \cite{vaswani2017attention}, in the present unsupervised speech enhancement framework. The DVAE models may be further boosted with more expressive latent variables, e.g., introducing hierarchical multi-scale structure and normalizing flows \cite{NEURIPS2020_e3b21256}. We also plan to extend the proposed method to a multi-modal framework, using the speaker's lips motion and visual appearance, in the continuation of VAE-based audio-visual speech enhancement \cite{sadeghi2020audio}.





\ifCLASSOPTIONcaptionsoff
  \newpage
\fi



\bibliographystyle{IEEEtran}
\bibliography{bibtex/taslp}

%



%

\begin{IEEEbiography}[{\includegraphics[width=1in,height=1.25in,clip,keepaspectratio]{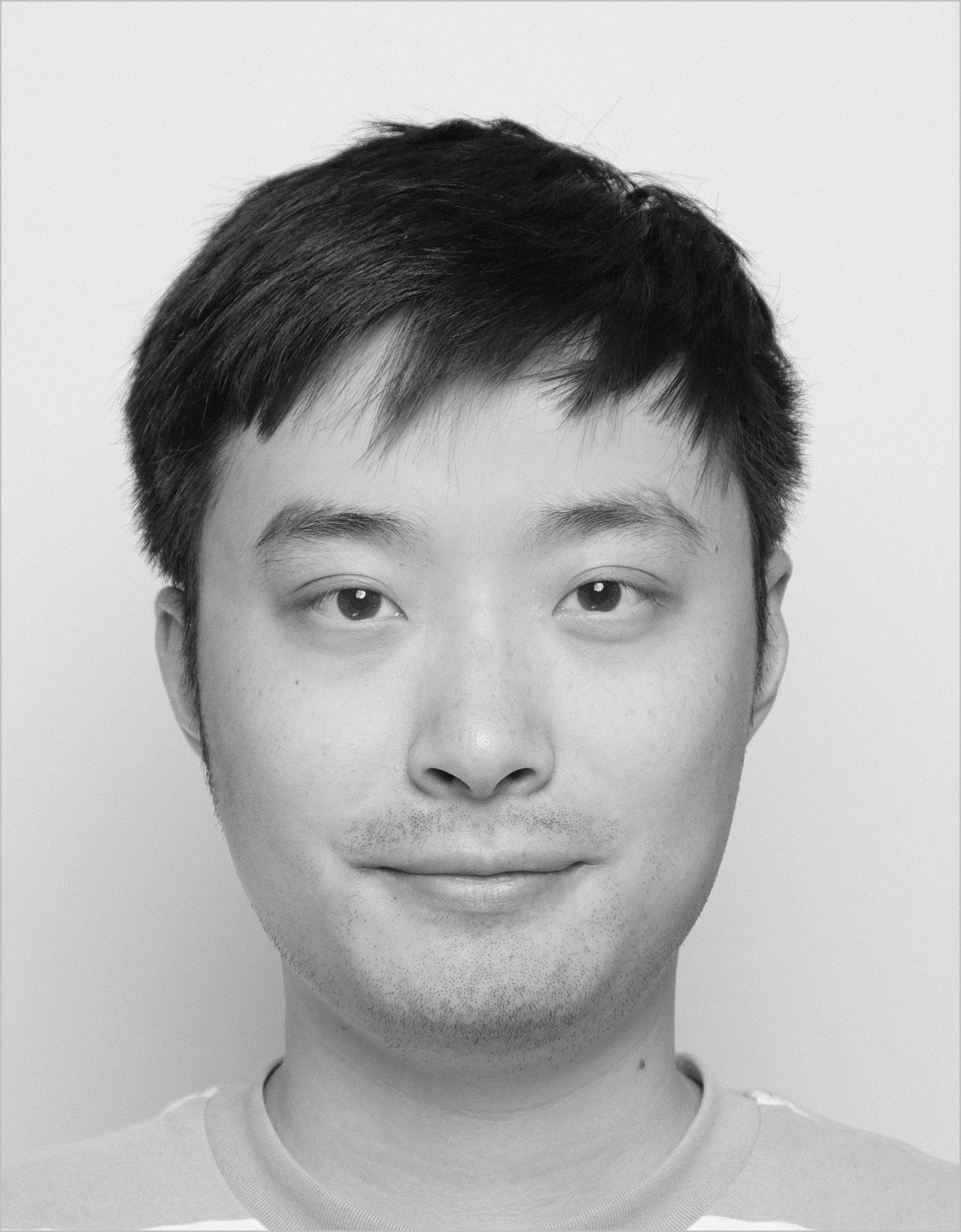}}]{Xiaoyu BIE}
received the B.Sc. degree in optical and electronic information from Huazhong University of Science and Technology, Wuhan, China, in 2016, the Engineering degree in applied optics from Institut d’Optique/University of Paris-Saclay, Gif-sur-Yvette, France, in 2018 and the M.Sc. degree in signal and image processing from CentraleSup\'elec/University of Paris-Saclay, Gif-sur-Yvette, France, in 2018. He is currently working toward the Ph.D. degree with INRIA and Univ. Grenoble Alpes. His research focuses on deep sequential generative model, speech analysis and human understanding.
\end{IEEEbiography}

\begin{IEEEbiography}[{\includegraphics[width=1in,height=1.25in,clip,keepaspectratio]{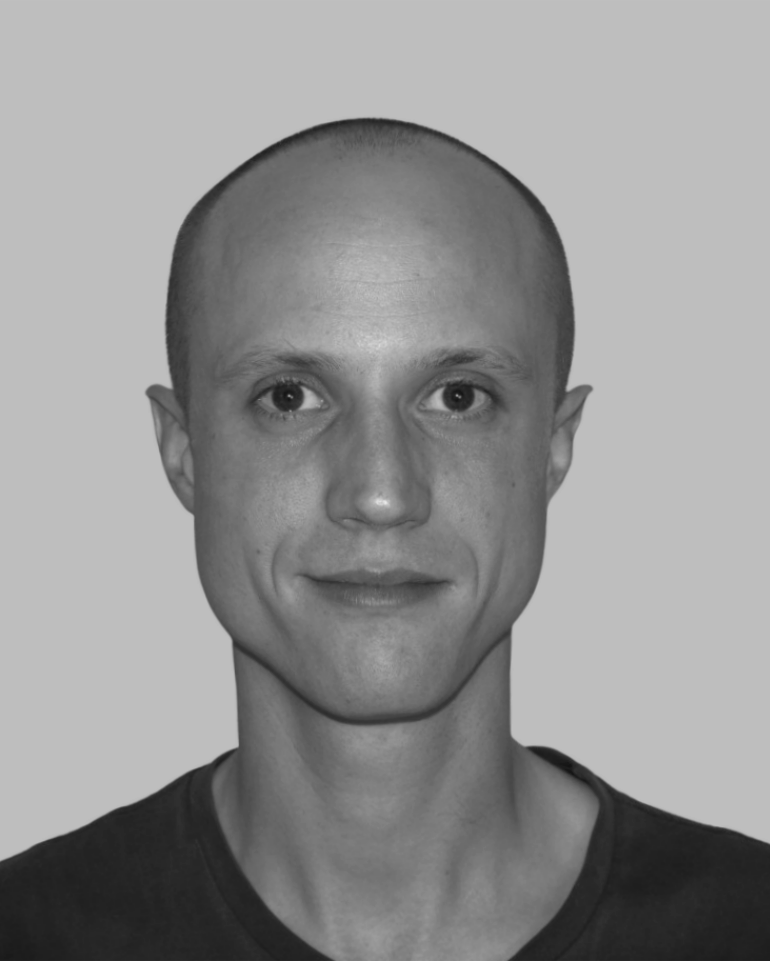}}]{Simon Leglaive} is an Assistant Professor (tenured) at CentraleSupélec and a researcher in the AIMAC team of the IETR laboratory, a CNRS joint research unit in Rennes, France. He received the Engineering degree from Télécom Paris (Paris, France) and the M.Sc. degree in acoustics, signal processing and computer science applied to music (ATIAM) from Sorbonne University (Paris, France) in 2014. He obtained the Ph.D. degree from Télécom Paris in the field of audio signal processing in 2017. He was then a post-doctoral researcher at Inria Grenoble Rhône-Alpes (Grenoble, France), in the Perception team.  His research focuses on signal processing and machine learning for audio and speech applications. He is mainly interested in weakly-supervised approaches for problems that consist in estimating latent signals from noisy and/or incomplete observations (e.g., source separation, speech enhancement).
\end{IEEEbiography}

\begin{IEEEbiography}[{\includegraphics[width=1in,height=1.25in,clip,keepaspectratio]{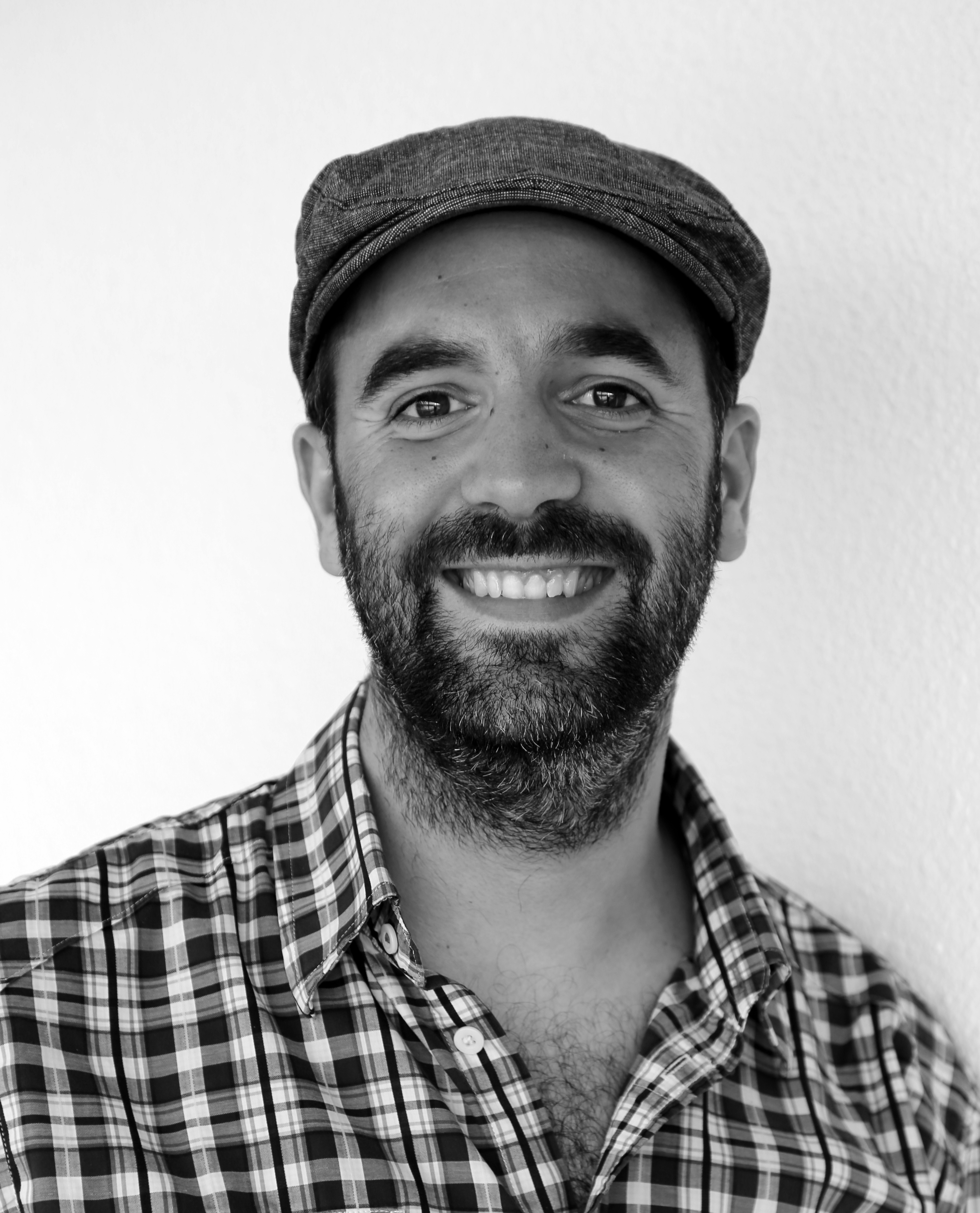}}]{Xavier Alameda-Pineda}
is a (tenured) Research Scientist at Inria, and the Leader of the RobotLearn Team. He obtained the M.Sc. (equivalent) in Mathematics in 2008, in Telecommunications in 2009 from BarcelonaTech and in Computer Science in 2010 from Université Grenoble-Alpes (UGA). He then worked towards his Ph.D. in Mathematics and Computer Science, and obtained it 2013, from UGA. After a two-year post-doc period at the Multimodal Human Understanding Group, at University of Trento, he was appointed with his current position. Xavier is an active member of SIGMM, a senior member of IEEE and a member of ELLIS. He is the Coordinator of the H2020 Project SPRING: Socially Pertinent Robots in Gerontological Healthcare and is co-leading the “Audio-visual machine perception and interaction for companion robots” chair of the Multidisciplinary Institute of Artificial Intelligence. Xavier’s research interests are at the cross-roads of machine learning, computer vision and audio processing for scene and behavior analysis and human-robot interaction.
\end{IEEEbiography}

\begin{IEEEbiography}[{\includegraphics[width=1in,height=1.25in,clip,keepaspectratio]{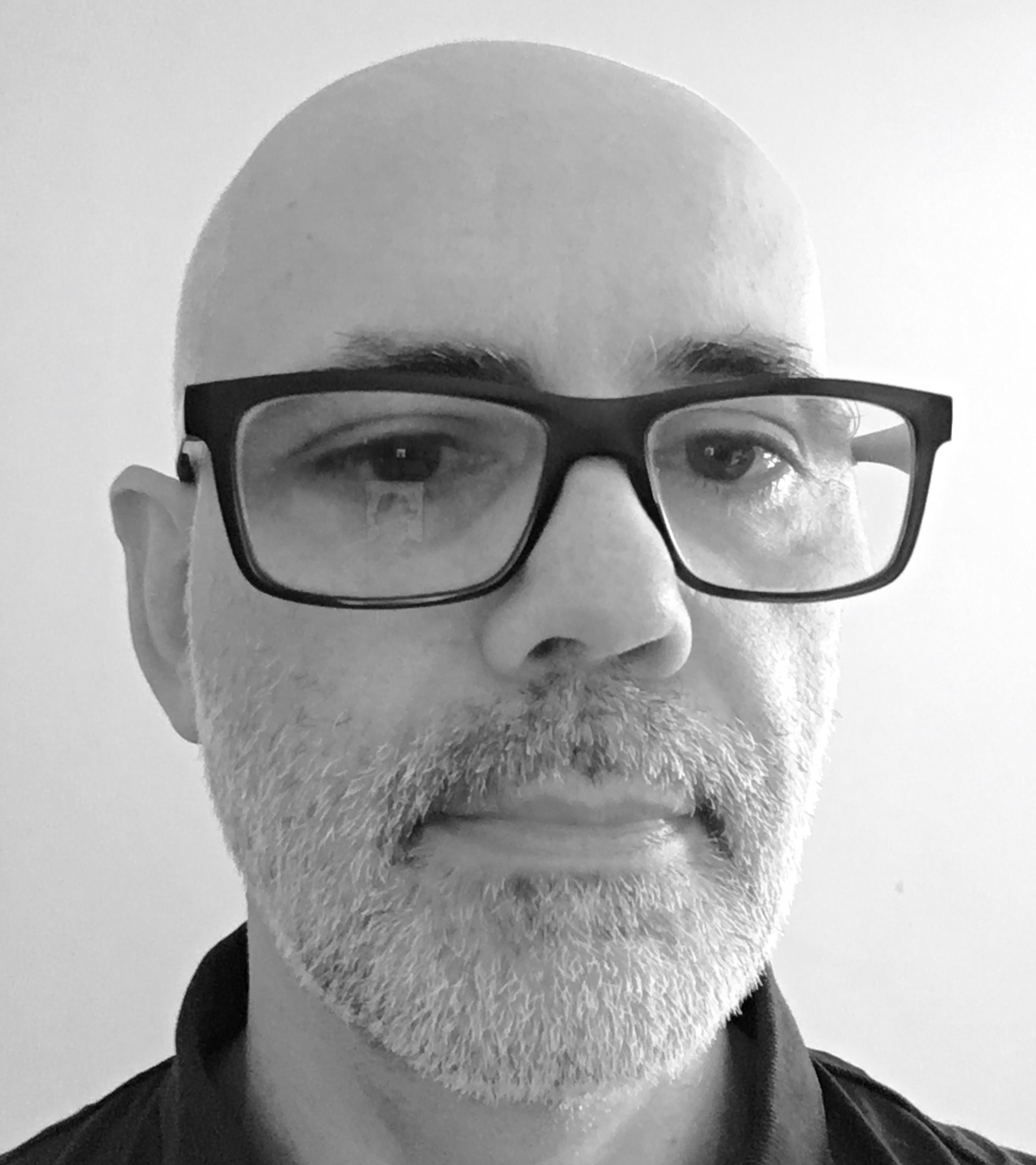}}]{Laurent Girin} received the M.Sc. (1994) and Ph.D. (1997) degrees in signal processing from Institut National Polytechnique de Grenoble (INPG), France. In 1999 he joined Ecole Nationale Sup{\'e}rieure d’Electronique et de Radio{\'e}lectricit{\'e} de Grenoble, as an Associate Professor. Currently he is a Full Professor at Grenoble Institute of Technology (Grenoble-INP) in the Physics, Electronics, and Materials (PHELMA) department, where he lectures signal processing theory and applications to audio. His research activity is carried out at GIPSA-Lab (Grenoble Laboratory of Image, Speech, Signal, and Automation). It deals with speech and audio processing (analysis, modeling, coding, transformation, synthesis, localization, enhancement and separation), with a special interest in multimodal speech processing (e.g., audiovisual processing or articulatory-acoustic modeling). Prof. Girin is currently the Head of the Speech and Cognition Pole at GIPSA-lab (one of the four scientific departments of GIPSA-lab). He is also a regular collaborator of the RobotLearn team at Inria Grenoble.
\end{IEEEbiography}








\end{document}